\documentclass[12pt,a4paper]{article}

\usepackage{dsfont}

\RequirePackage[OT1]{fontenc}
\RequirePackage{amsthm,amsmath}
\RequirePackage{natbib}
\RequirePackage[colorlinks,citecolor=blue,urlcolor=blue]{hyperref}
\RequirePackage{subfigure}
\RequirePackage{graphicx} 
\usepackage{float}
\usepackage{enumitem}
\usepackage{xcolor}
\usepackage{pdflscape}
\usepackage{amsfonts}

\usepackage{hyperref}

\makeatletter 

\makeatletter
\newcommand{\otherlabel}[2]{\protected@edef\@currentlabel{#2}\label{#1}}
\makeatother

\begin{document}

\title{\sc{Bayesian hierarchical modelling of sparse count processes in retail analytics}}
\author{James Pitkin, Ioanna Manolopoulou and Gordon Ross }

\maketitle

\begin{abstract}
The field of retail analytics has been transformed by the availability of rich data which can be used to perform tasks such as demand forecasting and inventory management. However, one task which has proved more challenging is the forecasting of demand for products which exhibit very few sales. The sparsity of the resulting data limits the degree to which traditional analytics can be deployed. To combat this, we represent sales data as a structured sparse multivariate point process which allows for features such as auto-correlation, cross-correlation, and temporal clustering, known to be present in sparse sales data. We introduce a Bayesian point process model to capture these phenomena, which includes a hurdle component to cope with sparsity and an exciting component to cope with temporal clustering within and across products. We then cast this model within a Bayesian hierarchical framework, to allow the borrowing of information across different products, which is key in addressing the data sparsity per product. We conduct a detailed analysis using real sales data to show that this model outperforms existing methods in terms of predictive power and we discuss the interpretation of the inference.
\end{abstract}

\section{Introduction}\label{SMG:Intro}
One of the main objectives of retail analytics is to build predictive demand forecasting models, for purposes such as inventory management, profit forecasting, assessing the impact of marketing to name but a few.
Demand models have been extensively studied in the literature, focusing on forecasting sales of high volumes \citep*{seeger2016bayesian, ferreira2015analytics, sahu2014hierarchical}. However, these forecasting models often struggle to capture the demand dynamics of products with low sales volumes. Such products, known as slow-moving-inventory (SMI), are typically for sale the entire year but are only purchased 1-5\% of days, often with an intermittent pattern. They are usually nonfood merchandise such as technology, fashion and general household items. The resultant demand data of SMI take the form of a sparse count process per product, largely populated with zeros, with auto-correlation and contemporaneous structure across different products (due to seasonality, promotions and current trends).

There are three main aspects of a predictive model of SMI which are challenging. Firstly, since these products have low sales volumes, this leads to an inflation of zeros (corresponding to days with no sales), which makes it difficult to learn the effect of traditional variables used in forecasting models (prices, promotions, seasonality). Secondly, SMI demand often occurs in bursts across different products, indicating a dependency either between a product's own sales history and the history of other similar products, or on a common external factor that cannot be accounted for by available covariates. 
Thirdly, SMI is often stocked and sold for a relatively limited amount of time (short sales cycles), which results in little covariate and demand history.

Previous research dealing with such zero-inflated bursty processes includes exponential smoothing and related methodologies that attempt to forecast future observations as a weighted moving average of past observations over time \citep*{croston1972forecasting,gardner2006exponential}. Such approaches primarily focused on the temporal burstiness of demand and demonstrated initial success, though lack an underlying stochastic process  consistent with intermittent demand and fail to provide a framework  that naturally accounts for predictors, information borrowing and uncertainty \citep*{shenstone2005stochastic}. More recent developments have included neural network approaches that show promise at finding the complex non-linear interdependencies across multiple  intermittent demand series across but suffer from over-fitting issues and lack an underlying interpretability \citep*{kourentzes2013intermittent, pour2008hybrid, mishra2014intermittent}. The closest approach striving to accommodate the zero-inflation, demand clustering and information sparsity exhibited in intermittent demand comes from  \cite{chapados2014effective}, who implement a Bayesian hierarchical zero-inflated count model with time-varying regression parameters that shares information across intermittent demand series. However, their approach limits the dependency on historical demand to an $AR(1)$ process in the mean of the count distribution and ignores the zero-process altogether, exclude pricing information from their framework and without considering contemporaneous dependence between intermittent demand series. Though existing approaches have demonstrated a degree of success at forecasting the intermittent demand of SMI, none have developed a unified model that incorporates excitation dynamics, covariates beyond just seasonality and information pooling between the intermittent demand series in a way that sheds light into additive benefits that each of these components has with respect to forecasting performance.

In this work, we develop modelling, inferential and predictive methods able to learn the dynamics of sparse count processes for SMI products with few to no sales. We flexibly introduce covariates into the self-exciting model for sparse processes of  \cite{porter2012self}. We extend the model to include a cross-excitation contribution that allows differing intermittent demand series to excite one another, capturing the process of intertwined contemporaneous excitation dynamics observed in SMI data.  We overcome the lack of information for each product by integrating individual products into a Bayesian hierarchical model that accommodates shrinkage and information passing across differing sparse count process, without requiring the data for each product to exist over the same time period. 

The layout of this paper is as follows; section \ref{SMG:Data} describes the SMI demand data used in this paper. Section \ref{SMG:Background} describes hurdle models and the Hawkes process. Section \ref{SMG:Model} outlines our hierarchical Bayesian hurdle model with self and cross-excitation components to model multiple sparse count processes simultaneously. Section \ref{SMG:Results} presents the results of our sparse count process on the demand data of touchscreen tablets across five South London supermarkets. We conduct a detailed investigation to compare our model to its non-hierarchical equivalent and models without the self and cross-excitation terms to highlight the benefits of the information borrowing and excitation components and discuss the implications of these results within the context of retail analytics. Section \ref{SMG:Conclusion} concludes with a summary of our contributions and a discussion of possible future developments.

\section{Data}\label{SMG:Data}
We implemented our methods on a dataset recorded through electronic points of sale of a leading UK supermarket retailer, anonymised for general research purposes and that no individual shoppers could be identified. Access to the anonymised dataset was provided by dunnhumby ltd. The data consist of 17 longitudinal SMI sales processes over 464 days of trading between the dates $1^{st}$ October 2013 to $7^{th}$ January 2015. For each product, the daily count corresponds to the aggregated sales of a touchscreen tablet across five large supermarkets within south London. Daily prices as well as seasonality characteristics are available as covariates during the 464 trading days, during which all of the 17 tablets were stocked and in circulation.  We split the data into training and test sets, the first 364 trading days between $1^{st}$ October 2013 to $29^{th}$ September 2014 (a full trading year excluding Christmas), and the remaining 100 trading days between $30^{th}$ September 2014 to $7^{th}$ January 2015 kept as hold out test set. These training and test split gives a balance between providing sufficient training periods where we observe one full year to allow the learning of seasonal trends, whilst having test sets of a reasonable size to allow meaningful forecasts. This dataset is challenging since we only have one year to learn seasonality from and thus makes a hierarchical model formulation particularly applicable.


Table \ref{table:sale_rates} provides summary statistics over the training set of the sale counts across the 17 tablet products. The demand across the category is primarily driven by one product, as it accounts for 75\% of sales. However, the remaining products are extremely slow moving as indicated by the majority of them only having  0.5-5\% non-zero sales days.

\begin{table}[H]
\caption{Summary statistics of SMI demand within tablet category on the training set. The brands have been anonymised with fictitious names for privacy purposes.}\label{table:sale_rates}
\label{sphericcase}
\centering
\begin{tabular}{cccc}
\hline
Product & Brand  & total sales &   \% non-zero sale days  \\
\hline
1 & SPARK & 1 & 0.27 \\ 
  2 & TECHY & 409 & 53.57 \\ 
  3 & TECHY & 36 & 4.12 \\ 
  4 & GADGET & 9 & 1.92 \\ 
  5 & TECHY & 5 & 1.37 \\ 
  6 & TECHY & 13 & 3.57 \\ 
  7 & TECHY & 13 & 3.57 \\ 
  8 & GADGET & 13 & 3.30 \\ 
  9 & GADGET & 2 & 0.27 \\ 
  10 & GADGET & 5 & 1.37 \\ 
  11 & TECHY & 1 & 0.27 \\ 
  12 & TECHY & 12 & 1.92 \\ 
  13 & TECHY & 2 & 0.55 \\ 
  14 & TECHY & 3 & 0.82 \\ 
  15 & TECHY & 9 & 0.82 \\ 
  16 & TECHY & 6 & 1.10 \\ 
  17 & TECHY & 3 & 0.82 \\ 
   \hline
\end{tabular}
\end{table}

\begin{figure}[H]
  \centering
     \includegraphics[trim={0 0.5cm 0 1.8cm},clip,width=0.49\textwidth]{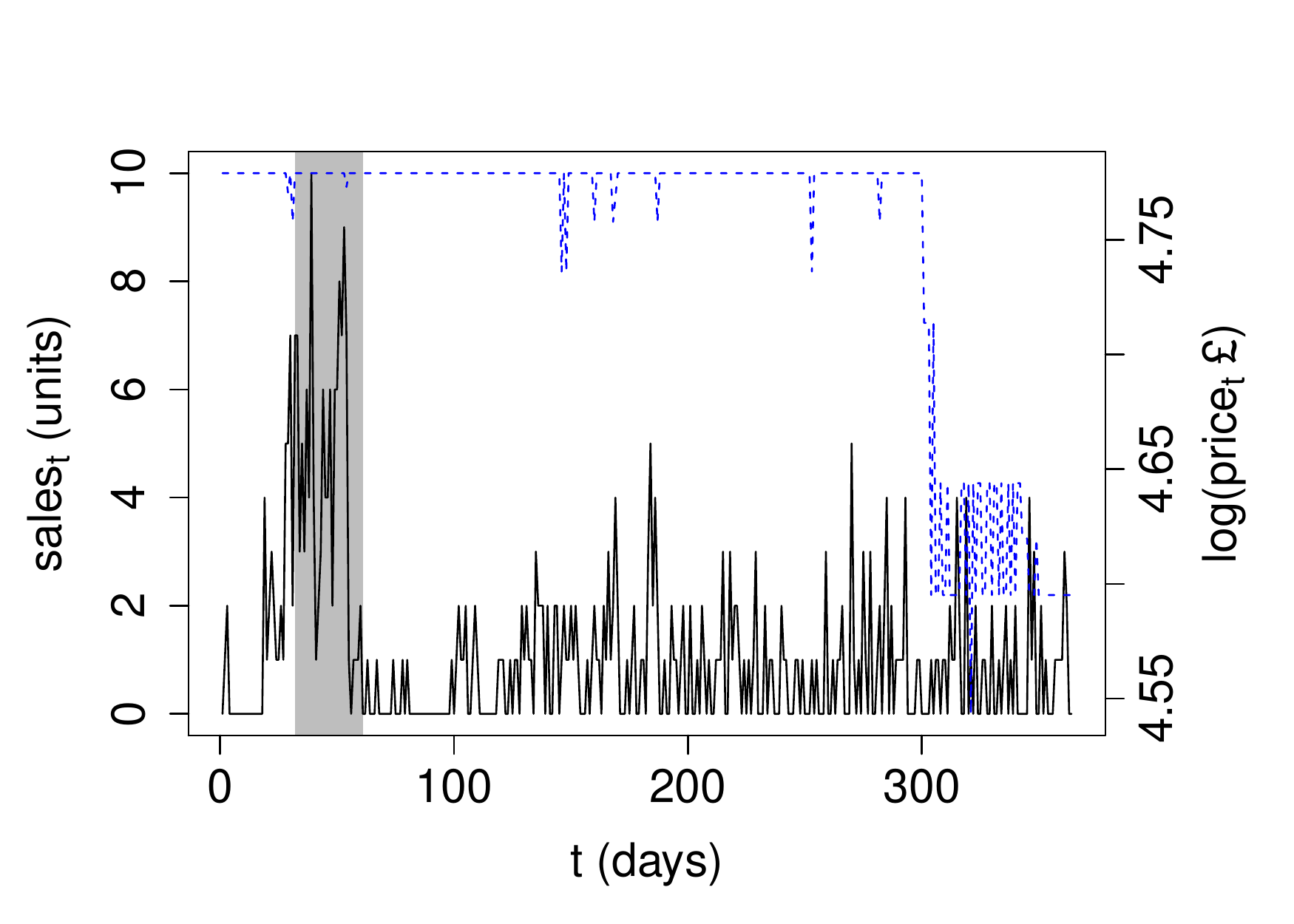}
     \includegraphics[trim={0 0.5cm 0 1.8cm},clip,width=0.49\textwidth]{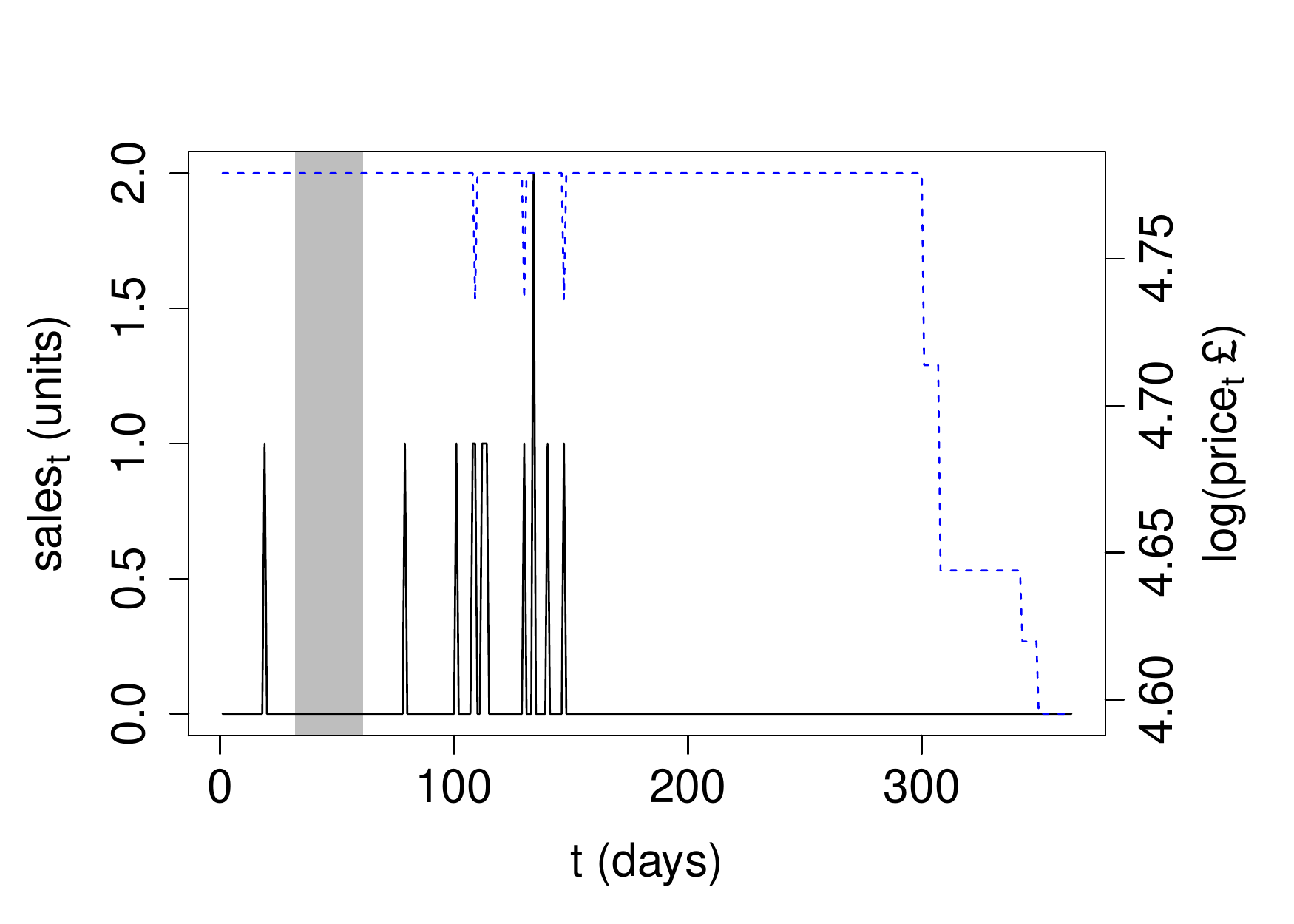} 
    \caption{\label{plot:demandvsprices}Plots of demand series (solid black line) for two tablets with their respective log prices in $\pounds$ (dashed blue) over 364 days of training data.  The left panel is a high volume tablet and the right panel is the demand of a low volume tablet. The shaded region is the month prior to Christmas.}
\end{figure}

\begin{figure}[H]
  \centering
     \includegraphics[trim={0 0.5cm 0 1.8cm},clip,width=0.49\textwidth]{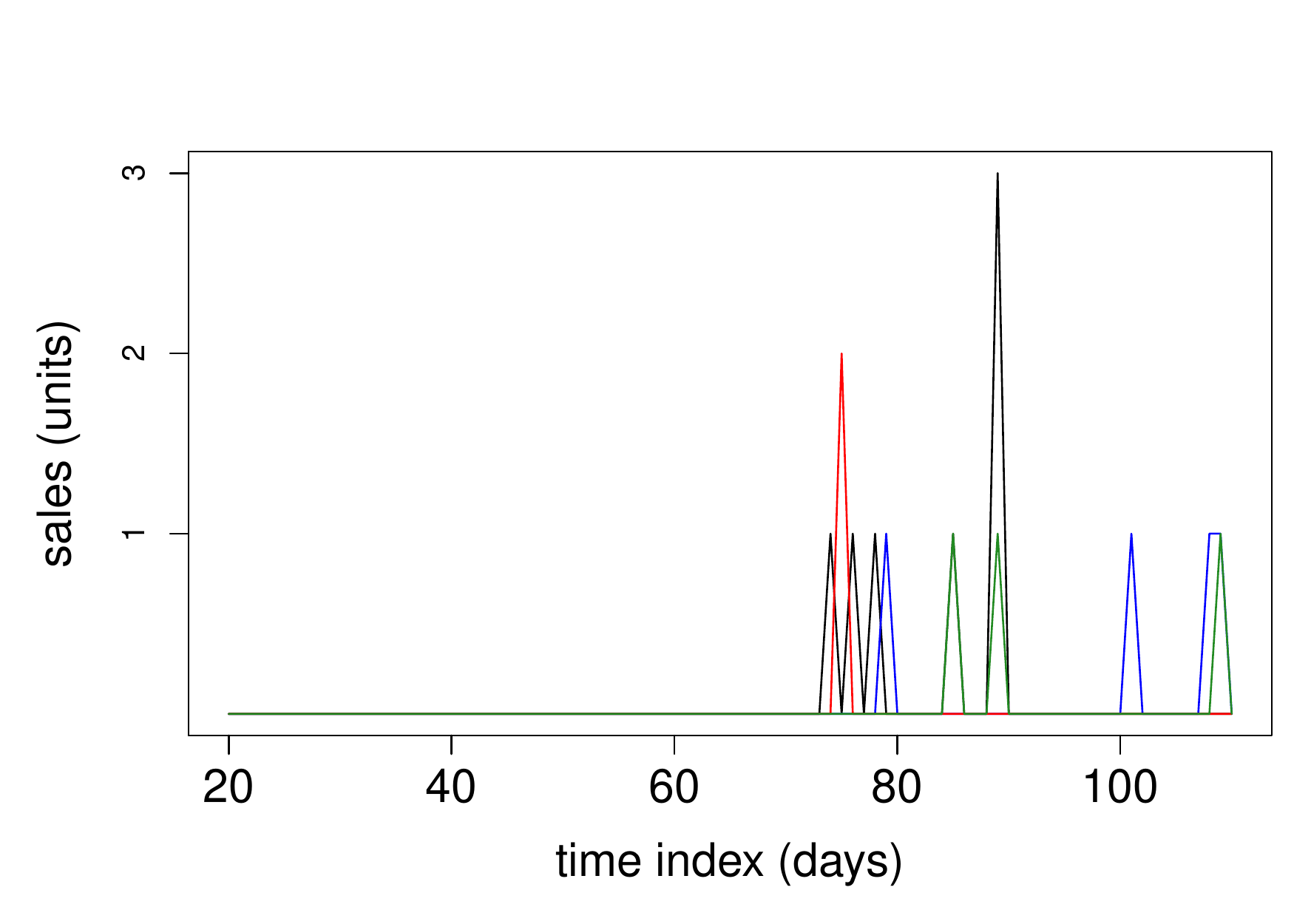}
     \includegraphics[trim={0 0.5cm 0 1.8cm},clip,width=0.49\textwidth]{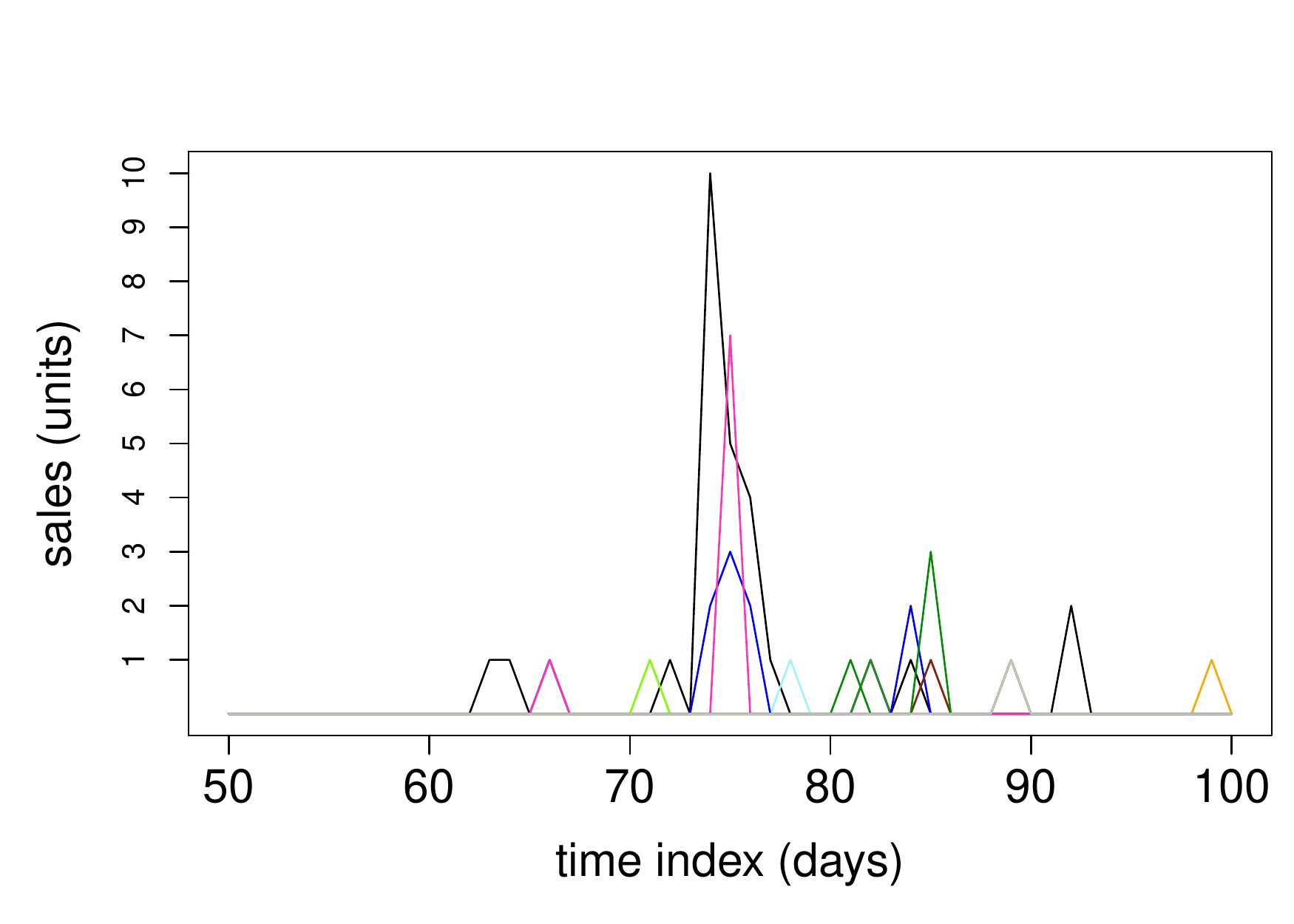} 
    \caption{\label{plot:cross_exciation}Plots of tablet sales across two brands over portions of the training set. The left plot corresponds to the GADGET brand and the right plot to the TECHY brand. For each of the plots, the differing colours correspond to the sales of a particular product within the given brand. }
\end{figure}

These data demonstrate many of the pertinent features of SMI sales processes. Figure \ref{plot:demandvsprices} contrasts the sales and respective prices of one of the faster-selling tablets against a slower one. The plots illustrate the zero-inflation, especially in relation to the length of the observed time period and that the sales do not show a straightforward dependence on either the prices or the seasonal effects, as indicated by the little movement in demand with respect to changes in prices and season. A clustering effect in the succession of sales within their own demand series is also evident. For example, sales of the right-hand plot in Figure \ref{plot:demandvsprices} fall during the month prior to the festive period, typically thought of as driving demand, but a quick succession of sales follows shortly after this month. This suggests an excitation process not accounted for by covariate information, as sales bursts occur outside the effects explained by covariate data. Figure \ref{plot:cross_exciation} provides plots suggesting the existence of possible contemporaneous excitation of tablet sales within a particular brand. We see that sales of a tablet in a given brand are often followed by a subsequent sale of another tablet of the same brand.

\section{Background}\label{SMG:Background}

Our aim is to develop a Bayesian hierarchical model for the sales of product $i$ on day $t$, denoted $y_{it}  \in \{0,1, \ldots \}$. We will decompose the model into a `zero' (days with no sales) and `non-zero' (days with non-zero sales) using a hurdle model to capture the zero-inflation in the count processes, and will combine this with self- and cross-excitation components to account for the clustering of events. To this end, we now review two main approaches used to handle the inflation of zeroes in the sales process and the apparent excitation, namely hurdle regression models to deal with the abundance of zeros exhibited in the count process and shot noise processes to handle the dependency of  sales on their immediate history.

\subsection{Hurdle models}
\cite{mullahy1986specification} introduced the hurdle regression model to handle an inflation of zeros in count data that traditional count models (Poisson, negative binomial regression) could not adequately account for. The hurdle model defines a distribution over the counts $\{0,1,\ldots \} $ and assumes these counts can be split into two separate processes; a process accounting exclusively for the 0's (the hurdle), and a process accounting for non-zero counts. 
Hurdle models, unlike their zero-inflated model counterpart \citep{lambert1992zero}, assumes the zero and non-zero processes are separable, as 0 observations arise exclusively from the degenerate 0 distribution and the count distribution over $\{1,\ldots \} $.  We opt for a hurdle model over a zero-inflated model due to the separability of the zero and count processes (that accommodates efficient inference) and so that any occurrences of 0 can be directly linked to the zero process. 

Within our context of SMI modelling, the inflation of zeros corresponds to days when we observe zero sales, and the count process corresponds to days when we observe non-zero sales. More concretely, given $y_{t}$ sales, the probability density function of the hurdle model given covariates  $\boldsymbol{x}_t$ can be specified as:
    \begin{equation}\label{ZI_model}
  p(y_t \mid \boldsymbol{x}_t, \boldsymbol{\theta})= \left\{
     \begin{array}{lr}
 p(\boldsymbol{x}^z_t ,  \boldsymbol{\theta}^z)  \text{, for } y_t=0\\
         (1- p(\boldsymbol{x}^z_t ,  \boldsymbol{\theta}^z))   f(y_t \mid \boldsymbol{x}^c_t, \boldsymbol{\theta}^c) \text{, } y_t=1,\ldots 
              \end{array}
   \right. 
\end{equation} 
Here  $ p(\boldsymbol{x}^z_t, \boldsymbol{\theta}^z)$ is the probability of observing a zero count at time $t$ and $f( \cdot \mid \boldsymbol{x}^c_t, \boldsymbol{\theta}^c)$ is a probability mass function defined on the positive integers.  The covariates for the zero  process $\boldsymbol{x}^z_t$ and count process $\boldsymbol{x}^c_t$ may overlap. The $\boldsymbol{\theta}^z, \boldsymbol{\theta}^c$ are parameters for the zero and count processes respectively. For notational purposes, we let $E_{t}$ be the indicator for an event day such that $E_{t} =1$ if $y_{t} \geq1$ (a day $t$ where at least one sales instance is observed) and $E_{t} =0$ if $y_{t} = 0$ (a day $t$ with no sales).

\subsection{Self-exciting processes}\label{subsec:self_excitation}
\begin{figure}[t]
  \centering
     \includegraphics[trim = {0 0.5cm 0 1.8cm}, clip, width=0.8\textwidth]{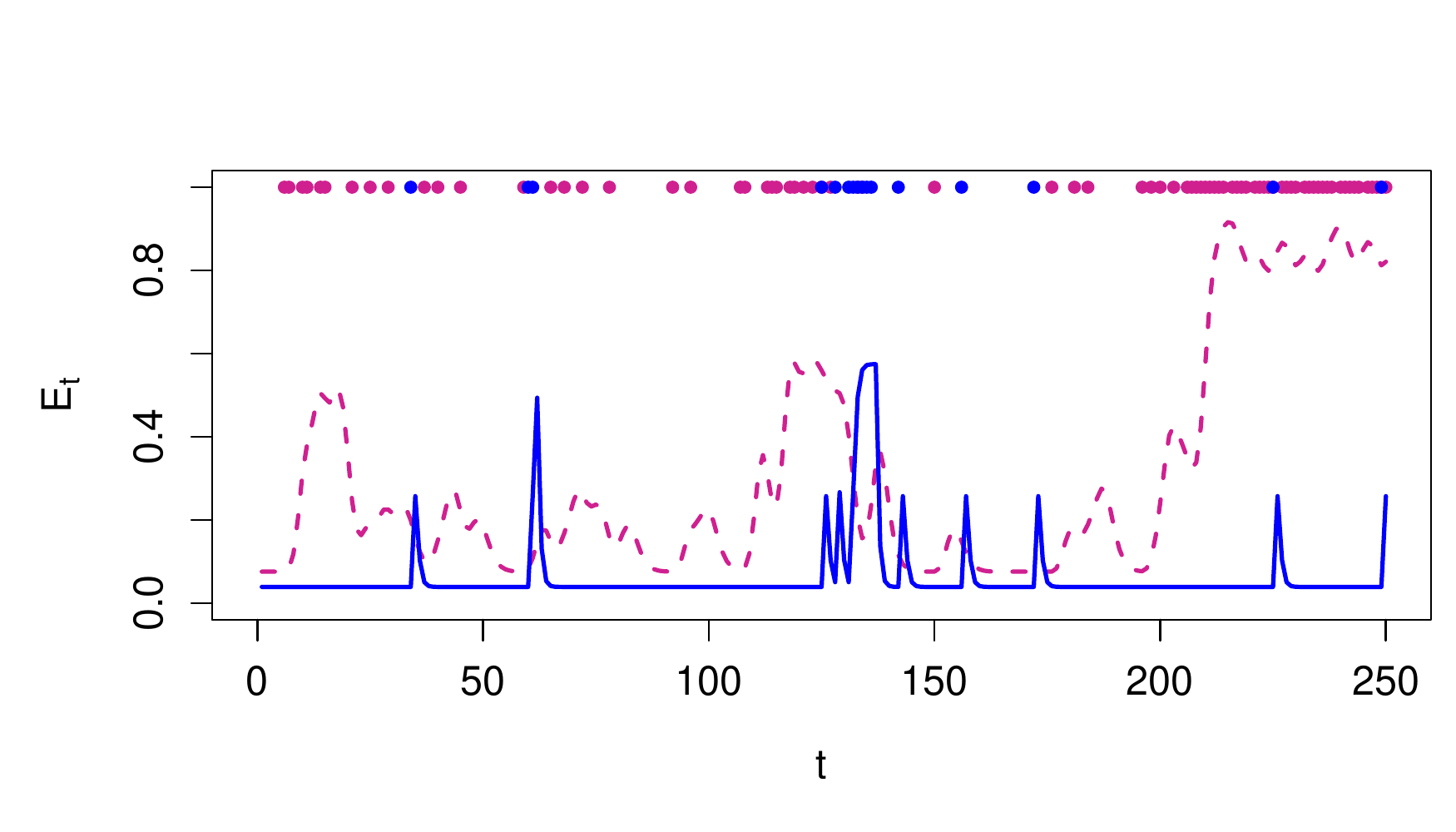}
    \caption{\label{HP_logistic_simulate} Simulated example. Two series of samples are generated from $E_{t} \sim \text{Bernoulli}(p_{t})$, with $\text{logit}(p_{t}) = \theta+ \kappa\sum_{i < t}E_{i}g(t-i \mid \mu, \tau) $ for $t=1,\ldots,364$ where $g( \cdot \mid \mu, \tau)$ is the truncated negative binomial density on the positive integers with mean and scale $\mu, \tau$. The blue dots are $E_t$ samples generated from $ \left( \theta, \kappa, \mu, \tau \right) = \left( -3.2, 3.1, 1.0, 5.0 \right)$ and the solid blue line is the corresponding $p_t$. The maroon dots are $E_t$ samples generated from $ \left( \theta, \kappa, \mu, \tau \right) = \left( -2.5, 5, 5, 60 \right)$ and the dashed maroon line is the corresponding $p_t$. We observe how the differing $ \left( \theta, \kappa, \mu, \tau \right) $ lead to different clustering patterns and underlying shapes of the probability of events.    
    }\label{plot:brandcross}
\end{figure}

\cite{hawkes1971spectra} introduced a Hawkes process as a self-exciting temporal point process with conditional intensity function
    \begin{equation}\label{conditional_intensity_continuous}
 \lambda(t) = \varphi(t) +\sum_{i:t_{i}<t} \nu(t-t_i)
    \end{equation}
where $\varphi(t)$ is the background rate, $t_{i}$ are the times prior to time $t$ when an event (i.e.~non-zero sales) occurred and $\nu(\cdot)$ a continuous excitation function that controls the extent to which events cluster together. This process effectively describes a count process where events increase the probability of further such events in the short term, leading to clustered events (in our case, days with non-zero sales). In the discrete context, the above can be re-expressed as:
    \begin{equation}\label{conditional_intensity}
 \lambda(t) = \varphi(t) +\sum_{j<t} \kappa E_{j}g(t-j)
    \end{equation}
where $\varphi(t)$ is, as before, the background rate, $E_{t}$ is a Boolean indicator indicating event days ($E_{t} =1$ for an event day, i.e.~a day with non-zero sales), $g(\cdot) \geq 0$ is the excitation kernel (a probability mass function) that controls the extent to which events cluster together and $\kappa$ is some trigger constant that can be interpreted as the average number of triggered events produced by each event. With a Hawkes process, instances of an event in turn increase ($\kappa >0$) or decrease ($\kappa <0$) the probability of further such events occurring in the future. In this work we focus on the case $\kappa>0$ which represents excitation (rather than inhibition). We denote the history of events up to but not including $t$ as $H_{t-1} = \left(E_{1}, \ldots, E_{t-1}    \right)$. Figure \ref{HP_logistic_simulate} plots two simulated series from a Bernoulli distribution with a Hawkes process term. It illustrates the variation in Bernoulli samples depending on the parameters of the excitation kernel and trigger constant. For example, the maroon curve with the higher excitation constant $\kappa$ shows much stronger excitation as exhibited by the densely clustered events dots, as opposed to the blue which are mostly isolated events.

\subsection{Cross-exciting processes}
Various extensions to (\ref{conditional_intensity}) have been made to include cross excitation across related spatial or temporal processes. \cite{lai2014topic} proposed a scheme allowing for inter-excitation and inhibition across different social media events across both time and space domain. They used a triggering kernel specified as exponential in time and Gaussian in space to capture cross excitation and inhibition in tweets in different topics and geographies.  \cite{zhou2013learning} used multi-dimensional Hawkes process (in the continuous space) to model information spread across sparse low-rank social networks and a triggering function which incorporates excitation from connected individuals in an additive form. \cite{blundell2012modelling} modelled interaction between human relationships using linked Hawkes processes through a kernel trigger function for the cross entries, which are linked via a non-parametric Chinese restaurant process to determine the partitions amongst social groups.  Although the aforementioned approaches demonstrate a degree of success within their relevant contexts, they have not been applied to sales forecasting before. In addition, multivariate Hawkes processes require specifying excitation relationships between all events pairs of the multivariate point process, which increases model  complexity and can be computationally challenging to infer.

\section{Model}\label{SMG:Model}
We model the daily sales of SMI by explicitly modelling the absence of a sale (termed the `zero-process'), and the number of sales by the `count-process'. Our model uses a Bayesian hierarchical version of the hurdle model of (\ref{ZI_model}) with self and cross-excitation terms discussed in section \ref{subsec:cross_excitation} in both the zero and count components. Our proposed model makes the following three extensions to existing models; firstly we use covariates beyond seasonal information, in particular we use price along boolean seasonal variables to assist in forecasting sales. Secondly, we use cross-excitation in the zero process of (\ref{ZI_model}) that aims to capture the contemporaneous nature of sales bursts across the SMI category. Thirdly, we build a Bayesian hierarchical model across the sales $y_{it}$ (the sales at product $i$ at time $t$) of a SMI category to allow information borrowing which is key in addressing the sales sparsity per product. 


\subsection{Covariate data}\label{SMI:covariates}
In addition to the excitation exhibited in SMI sales, product level covariates may offer predictive power to SMI forecasting. We introduce covariate data into the model through the background intensity function $\varphi(t)$ of (\ref{conditional_intensity}). 
In the supermarket sales context, this corresponds to a product's own price along with seasonal effects (which are common for all products). In particular, these covariates for a product $i$ at time $t$ are logarithm of its price, along with the indicator functions of week day, month and Christmas period. We summarise these covariates as:
$$\log(\text{p}_{it}) =   \log({\text{price}_{it} }) = \text{logarithm price of SMI product $i$ at time $t$,}$$
$$\text{s}_t =    \left(\mathds{1}_{(t \in \text{Christmas})}, \mathds{1}_{(t \in  \text{Mon})}, \ldots,\mathds{1}_{(t \in  \text{Sat})},\mathds{1}_{(t \in  \text{Jan})},\ldots, \mathds{1}_{(t \in  \text{Nov})}\right).$$ 
Using boolean indicators allows for a natural interpretation in an information borrowing scheme, and further avoids any explicit aggregation across the SMI product data, allowing us to easily handle any issues relating to products coming in and out of circulation. We specify the background intensities $\varphi_{i}^{z}(t)$,  $\varphi_{i}^{c}(t)$ of the zero and count processes of (\ref{conditional_intensity}) as:
    \begin{equation}\label{conditional_intensity_event_zero}
\varphi_{i}^{z}(t) = \theta^z_{i1} +\theta^z_{i2}\log(\text{p}_{it})+\sum_{k=1}^{18} \theta^z_{i,k+2}\text{s}_{kt}
    \end{equation}
      \begin{equation}\label{conditional_intensity_event_count}
\varphi_{i}^{c}(t) = \theta^c_{i1} +\theta^c_{i2}\log(\text{p}_{it})
    \end{equation}
where $ \{ \theta^z_{i1}, \ldots,\theta^z_{i20} \}$ and $ \{ \theta^c_{i1},\theta^c_{i2} \}$ are the parameters associated with the zero and count processes respectively for product $i$. The $j$ index of $\theta^z_{ij}$ ranges from $1-20$ to include the 1 additive constant, 1 log price variable, 6 week day, 11 month and 1 Christmas indicators. Functions (\ref{conditional_intensity_event_zero}) and (\ref{conditional_intensity_event_count}) describe the background intensities of the processes absent of excitation. Thus, in the zero process, we expect the background intensity to depend on a linear combination of log(price), seasonal effects and some additive constant through  a given link function, whereas in the count process, we expect the background intensity to depend on a linear combination of log(price) and some additive constant through a given link function. We restrict the background intensity of the count process to exclude seasonal effects to reduce model complexity and the possibility of over-fitting. It is important to note that, for a given product, the count process only exists for $t$ with $E_{t} =1$. This reduces the count process data compared to the zero process. The link functions of  (\ref{conditional_intensity_event_zero}) and (\ref{conditional_intensity_event_count})  are context-specific and will be specified in the data analysis sections.  We now denote these covariates as $\boldsymbol{x}^z_{it} = \left(\text{p}_{it}, \text{s}_t  \right)     $ and $\boldsymbol{x}^c_{it} = (\text{p}_{it}) $ for the zero and count processes respectively in line with notation of (\ref{ZI_model}).

\subsection{Cross-excitation}\label{subsec:cross_excitation}
SMI sales of different but comparable products may occur in contemporaneous `bursts', in that sales of a particular product may be followed by sales of a comparable product in the immediate future; these bursts can be a result of external advertising campaigns or viral dynamics, but importantly the apparent excitation not only happens auto-correlatively, but also contemporaneously across products. In the SMI context, cross-excitation is suspected to occur within brand, i.e.~a sale for a product leads to a higher probability of a sale of a product from the same brand over the subsequent days. Concretely, we define $\tilde{E}_{it} $ as the indicator for a \it cross event day \rm of product $i$ of some brand such that $\tilde{E}_{it}=1$ if $\sum_{k \in B \setminus \{ i\}}  y_{kt} \geq 1$, where $B$ is the set of indices corresponding to products of the brand, and $\tilde{E}_{it}=0$ if  $\sum_{k \in B \setminus \{ i \}} y_{kt} = 0$. Thus the indicator $\tilde{E}_{it}$ is 1 if there is at least one sale within the brand at time $t$ and 0 otherwise. We denote the history of cross-events up to but not including $t$ as $\tilde{H}_{it-1} = \left(\tilde{E}_{i1}, \ldots, \tilde{E}_{it-1}    \right)$.

The corresponding shot noise process with the self and cross-excitation of product $i$ then becomes:
 \begin{equation}\label{self_shot}
S_{it} = \sum_{j<t} \kappa_{i} E_{it}g(t-j  \mid \zeta_i ) 
    \end{equation}
 \begin{equation}\label{cross_shot}
\tilde{S}_{it} = \sum_{j<t} \tilde{\kappa}_{i} \tilde{E}_{it} g(t-j \mid \tilde{\zeta}_i) 
    \end{equation}
where $\kappa_{i}, \tilde{\kappa}_{i}$ are the trigger constants for the self and cross-excitation respectively and $g$ is some probability mass function parametrised by $\zeta_{i}$ and  $ \tilde{\zeta}_{i}$ controlling the shape of future self and cross-excitation respectively. Our cross-excitation formulation of (\ref{cross_shot}) is closely related to the multivariate Hawkes process \citep{hawkes1971spectra}, where we fix all cross-excitation kernels of a given product to 0 that correspond to a different brand, and have shared cross-excitation kernels with shared parameters for products corresponding to the same brand. We denote these collections of self and cross-excitation parameters as $\gamma_{i}= \left( \kappa_{i}, \zeta_{i} \right) $ and $\tilde{\gamma}_{i} = \left( \tilde{\kappa}_{i}, \tilde{\zeta}_{i} \right) $  respectively.

\subsection{Self and cross exciting hurdle model}\label{self_cross_hurdle_model}
We formulate our SMI model by utilising the hurdle model specification of (\ref{ZI_model}). In particular, we use a logistic link function to model the zero-process, with a background intensity  $\varphi^{z}(t)$  (\ref{conditional_intensity_event_zero})  including seasonal boolean covariates, logarithm of price, as well as  self and cross-excitation components ((\ref{self_shot}) and (\ref{cross_shot})). Similarly, for the count process we use a Negative Binomial distribution with a log-link mean intensity $\varphi^{c}(t)$ (\ref{conditional_intensity_event_count}) which includes logarithm of price as well as the self excitation term of (\ref{self_shot}). Our model is indexed by 17 longitudinal sales series from the tablets category over 464 (training+test) days of trading between the dates $1^{st}$ October 2013 to $7^{th}$ January 2015. We specify the probability mass function of the hurdle model as:
    \begin{equation}\label{SMI_model}
  p(y_{it} \mid \boldsymbol{x}_{it}, H_{it}, \tilde{H}_{it}, \boldsymbol{\theta}_{i})= \left\{
     \begin{array}{lr}
      p( \boldsymbol{x}^z_{it}, H_{it}, \tilde{H}_{it}, \boldsymbol{\theta}^z_i)  \text{, for } y_{it}=0\\
         (1- p(\boldsymbol{x}^z_{it},  H_{it}, \tilde{H}_{it}, \boldsymbol{\theta}^z_{i})   f(y_{it} \mid \lambda \left(\boldsymbol{x}^c_{it},H_{it}, \boldsymbol{\theta}^c_i\right), \phi) \text{, } y_{it} \in \mathbb{N}^{+}
     \end{array}
   \right. 
\end{equation}
where $\lambda(\cdot)$ represents a  link function and $f(y_{it} | \lambda, \phi)= \binom {y_{ik}-2+\phi}{y_{ik}-1} \left(\frac{\lambda-1} {\lambda-1+\phi}\right)^{y_{ik}-1}\left(\frac{\phi} {\lambda-1+\phi}\right)^{\phi} $ and $\phi=1$ which is the probability mass function of the shifted negative binomial distribution (NB) and $H_{it}$, $\tilde{H}_{it}$, $\boldsymbol{x}^z_{it}$ and $ \boldsymbol{x}^c_{i,t}$ are as defined in sections \ref{subsec:self_excitation}, \ref{subsec:cross_excitation} and \ref{SMI:covariates} respectively indexed by product $i$.  We specify the link functions as:

$$ \text{logit} \left( p\left(  \boldsymbol{x}^z_{it},H_{it}, \tilde{H}_{it},  \boldsymbol{\theta}^z_i\right) \right) = \varphi_{i}^{z}(t)+S_{it}^{z}+\tilde{S}_{it}^{z}$$ 
$$  \log(\lambda( \boldsymbol{x}^c_{it},H_{it}, \boldsymbol{\theta}^c_{i}) ) =  \varphi_{i}^{c}(t)+S_{it}^{c}$$  
$\varphi_{i}^{z}(t)$ and  $ \varphi_{i}^{c}(t)$ are as defined  from (\ref{conditional_intensity_event_zero}) and (\ref{conditional_intensity_event_count}) respectively but indexed by product $i$.   We define $S_{it}^{z} = \sum_{s<t} \kappa_{i}^{z} E_{it}g(t-s \mid \mu_{i}^{z}, \tau_{i}^{z} )$ and $\tilde{S}_{it}^{z} = \sum_{s<t} \tilde{\kappa}_{i}^{z} \tilde{E}_{it}g(t-s \mid \tilde{\mu}_{i}^{z}, \tilde{\tau}_{i}^{z} )$ similarly to (\ref{self_shot}) and (\ref{cross_shot}) respectively with $g(t \mid  \mu, \tau) = \binom {t-2+\tau}{t-1} \left(\frac{\mu-1} {\mu-1+\tau}\right)^{t-1}\left(\frac{\tau} {\mu-1+\tau}\right)^{\tau} $ as the shifted NB distribution. We similarly define $S_{it}^{c} = \sum_{s<t} \kappa_{i}^{c} E_{it}g(t-s \mid \mu_{i}^{c}, \tau_{i}^{c} )$. We denote the collection of shot parameters as $\boldsymbol{\tilde{\gamma}}_{i}^{z} = \left( \tilde{\kappa}_{i}^{z}, \tilde{\mu}_{i}^{z}, \tilde{\tau}_{i}^{z}  \right) $, $\boldsymbol{\gamma}^{z}_i = \left( \kappa_{i}^{z}, \mu_{i}^{z}, \tau_{i}^{z}  \right) $ and $ \boldsymbol{\gamma_{i}}^{c}  = \left( \kappa_{i}^{c}, \mu_{i}^{c}, \tau_{i}^{c}  \right) $ and collectively denote $\boldsymbol{\theta}^z_{i} = \left(\theta^z_{i1},\ldots,\theta^z_{i20}, \boldsymbol{\gamma_{i}}^{z}, \boldsymbol{\tilde{\gamma}}_{i}^{z} \right)$ and  $\boldsymbol{\theta}^c_{i} = \left(  \theta^c_{i1},  \theta^c_{i2}, \boldsymbol{\gamma_{i}}^{c} \right)$.

During this work, special attention is paid to the specification of hierarchical priors over the collection $\boldsymbol{\theta}^z_{i} $ and $\boldsymbol{\theta}^c_{i} $, as they are the mechanism through which we penalise complexity and pool information to combat data sparsity. In particular, we specify $ \theta^z_{ij} \sim \text{N}(	\rho_{j}^{z}, (\sigma_{j}^{z})^{2}) $ and $ \rho_{j}^{z} \sim \text{N}(\vartheta_{j}^{z}, (\zeta_{j}^{z})^{2}) $  and fix $({\sigma_{j}^{z})}^{2}$ for  $j=1,\dots,20$ and similarly  specify $ \theta^c_{ij} \sim \text{N}(\rho_{j}^{c}, (\sigma_{j}^{c})^{2}) $ and $ \rho_{j}^{c} \sim \text{N}(\vartheta_{j}^{c},(\zeta_j^{c})^{2}) $ and fix $(\sigma_{j}^{c})^{2} $  for each $j=1,2$.   For parameters of the shot function $S_{it}^{z}$, we specify $\gamma_{ij}^{z} \sim \text{Gamma}(\eta_{j}^{z}, \nu_{j}^{z}) $ with $\eta_{j}^{z} \sim \text{Gamma}(\alpha_{j}^{z}, \delta_{j}^{z}) $  and fix $\nu_{j}^{z}$  for each $j=1,2,3$. We  specify priors on  $ \tilde{\gamma}_{ij}^{z}$ and $\gamma_{ij}^{c}$ similarly. The full details of hierarchical prior specification are contained in appendix \ref{priors}.

\section{Results}\label{SMG:Results}
We fit variations of the model (\ref{SMI_model}) to the 17 longitudinal SMI sales processes over 364 days of trading between the dates $1^{st}$ October 2013 to $29^{th}$ September 2014. We denote time interval over which we train our models as $T^{\text{train}}$. A hold out test set over 100 trading days between $30^{th}$ September 2014 to $7^{th}$ January 2015 is used to evaluate the predictive performance of the model variations for both the zero and count processes. We denote this test interval as $T^{\text{test}}$.  As the zero and count processes are completely separable, we perform model inference and analysis separately.

\subsection{Zero process variations}\label{Zero_process_variations}
To assess the predictive benefits of the additions of self-excitation, cross-excitation and hierarchical components to the zero process of  the hurdle model of (\ref{ZI_model}), we implement the following cumulative variations of both the link functions as well as the hierarchical layering used in the modelling for each  $i=1,\ldots, 17$.

\begin{itemize}
\item \otherlabel{ZERO_NM1}{Base$_{1}^{z}$}  \bf Baseline model  (Base$_{1}^{z}$)\rm: We learn the zero process with link function $$ \text{logit} \left( p\left( \boldsymbol{x}^z_{it} ,H_{it}, \tilde{H}_{it},  \boldsymbol{\theta}^z_i\right) \right) = \varphi_{i}^{z},$$ 
 i.e.~a constant probability per product. This is the Bayesian baseline model as it estimates the zero-process independent of covariate information. The $\varphi_{i}^{z}$ is estimated using vague priors.  The performance of this model is used to verify the relative benefits that covariate information brings to SMI zero-process modelling.

\item \otherlabel{ZERO_HB}{HB$^{z}$}  \bf Hierarchical Bayesian  (HB$^{z}$)\rm: We learn the zero process  with link function $$ \text{logit} \left( p\left( \boldsymbol{x}^z_{it} ,H_{it}, \tilde{H}_{it},  \boldsymbol{\theta}^z_i\right) \right) = \varphi_{i}^{z}(t),$$ 
with the hierarchical prior formulation discussed in section \ref{self_cross_hurdle_model}. This model is implemented to establish a benchmark of the simplest regression model, i.e.~a model that excludes information of previous events and is used to verify the relative benefits of self excitation and cross-excitation.

\item \otherlabel{ZERO_BE}{BE$^{z}$}  \bf Bayesian with self-excitation  (BE$^{z}$)\rm: We learn the zero process of the hurdle model   with link function: $$ \text{logit} \left( p\left( \boldsymbol{x}^z_{it} , H_{it}, \tilde{H}_{it},  \boldsymbol{\theta}^z_i\right) \right) = \varphi_{i}^{z}(t)+S_{it}^{z},$$ 
but exclude the hierarchical prior formulation shown in section \ref{self_cross_hurdle_model}. More concretely, we fix the parameters $\rho_{j}^{z}, (\sigma_{j}^{z})^{2} $  and $\eta_{j}^{z}, {\nu}_{j}^{z}$  across all $j$.  This model is implemented to establish a  benchmark of a model with excitation but without information borrowing between products and is used to verify the relative benefits of information borrowing between products.

\item \otherlabel{ZERO_HBE}{HBE$^{z}$}  \bf Hierarchical Bayesian with self-excitation  (HBE$^{z}$)\rm:   We learn the zero process  with  link function $$ \text{logit} \left( p\left( \boldsymbol{x}^z_{it} ,H_{it}, \tilde{H}_{it},   \boldsymbol{\theta}^z_i\right) \right) = \varphi_{i}^{z}(t)+S_{it}^{z},$$ 
with the hierarchical prior formulation discussed in section \ref{self_cross_hurdle_model}.  This model is implemented to demonstrate the possible benefits of self-excitation in the standard zero inflated regression model. 

\item \otherlabel{ZERO_BEC}{BEC$^{z}$}  \bf Bayesian with self and cross-excitation  (BEC$^{z}$)\rm: We learn the zero process   with link function $$ \text{logit} \left( p\left( \boldsymbol{x}^z_{it} , H_{it}, \tilde{H}_{it},  \boldsymbol{\theta}^z_i\right) \right) = \varphi_{i}^{z}(t)+S_{it}^{z}+\tilde{S}_{it}^{z},$$ 
but exclude the hierarchical prior formulation shown in section \ref{self_cross_hurdle_model}. Prior specification is similar to that of \ref{ZERO_BE} but extended to include $ \boldsymbol{\tilde{\gamma}}_{i}^{z}$. This is a benchmark of a model with self and cross-excitation but without an information borrowing scheme. 

\item \otherlabel{ZERO_HBEC}{HBEC$^{z}$}  \bf  Hierarchical Bayesian with self and cross-excitation  (HBEC$^{z}$)\rm:  This  is the full model discussed in the section \ref{self_cross_hurdle_model}. We learn the zero process  with link function $$ \text{logit} \left( p\left(\boldsymbol{x}^z_{it} , H_{it}, \tilde{H}_{it},   \boldsymbol{\theta}^z_i\right) \right) = \varphi_{i}^{z}(t)+S_{it}^{z}+\tilde{S}_{it}^{z},$$ 
 with the hierarchical prior formulation discussed in section \ref{self_cross_hurdle_model}. The hyper-priors are selected to balance borrowing across products and penalising complexity. 

\end{itemize}

 Parameter inference is performed by Hamiltonian Monte Carlo sampling algorithm and is implemented using the \textsf{rstan} library \citep{RSTAN}.  Convergence was confirmed by Heidelberger Welch statistic  across all models and parameters \citep*{heidelberger1981spectral}.   The specification of hyper-priors is included in appendix \ref{priors}. For further MCMC implementation details, as well as additional model comparisons and discussion, refer to the supplementary materials.

\subsection{Zero process fits}\label{zero_fits}
The predictive performance of models \ref{ZERO_NM1},  \ref{ZERO_HB},   \ref{ZERO_BE},  \ref{ZERO_HBE},     \ref{ZERO_BEC} and \ref{ZERO_HBEC} is assessed by calculating how capable each model is at predicting the probability of a sale occurring on a given day over the test interval $T^{test}$  ($30^{th}$ September 2014 to $7^{th}$ January 2015) for each $i=1,\ldots, 17$ given the history of self  and cross events $H_{it}, \tilde{H}_{it}$, covariate information $ \boldsymbol{x}^z_{it}$ and posterior samples. We denote the $s^{th}$ posterior sample of $\boldsymbol{\theta}^z_{i} $ of the $i^{th}$ product as $\boldsymbol{\theta}^{z}_{is}$. The sales occurrence probabilities are based on the posterior samples $\boldsymbol{\theta}^{z}_{is}$ inferred from the training interval $T^{train}$ (between $1^{st}$ October 2013 to $29^{th}$ September 2014). More precisely, we apply the following methodology over the test interval:

\begin{enumerate}\label{test_method}

\item On given day $t$ on the test interval and $s^{th}$ posterior sample, we compute the full predictive posterior distribution of the probability of a sale occurring based conditioned on $  \boldsymbol{x}^z_{it},H_{it}, \tilde{H}_{it},  \boldsymbol{\theta}^{z}_{is}$ for each product $i=1,\ldots, 17$. 

\item We observe $y_{it+1}$ (the number of sales of product $i$ on day $t+1$) for each $i=1,\ldots, 17$ and update the self and cross event histories $H_{it+1}, \tilde{H}_{it+1}$ for $i=1,\ldots, 17$.

\item Repeat steps for each $t $, for each sample $s$ and $i$ over the test period of $30^{th}$ September 2014 to $7^{th}$ January 2015.

\end{enumerate}
This builds up a set of daily predictive posterior probabilities $p_{its}$ for each $s=1,\dots,S $ for the probability of a sale on a given day over $T^{test}$ for each $i=1,\ldots, 17$ based on posterior samples inferred from $T^{train} $ conditioned on $ \boldsymbol{x}^z_{it} ,H_{it}, \tilde{H}_{it},   \boldsymbol{\theta}^z_i$.
\newline
To evaluate the predictive performance of the models for the zero process we use the log posterior predictive density \citep*{gelman2014understanding}, denoted $\text{lppd}^{z}$,  given by:
$$\text{lppd}^{z}_{i} = \sum_{t \in T} \log\left( \frac{1}{S}\sum_{s=1}^{S}p_{its}^{E_{it}}(1-p_{its})^{(1-E_{it})}   \right)$$
where $p_{its}$ is the prediction probability of a sale occurring for product $i$ from posterior sample $s$ for some model of interest. Table \ref{table:zero_Gscores} provides the $\text{lppd}^{z} $ scores across products and models.

\begin{table*}
\centering
\caption{$\text{lppd}^{z,test}_{i}$ and $\text{lppd}^{z,train}_{i}$ scores of the zero process fits for the models \ref{ZERO_NM1}, \ref{ZERO_HB},  \ref{ZERO_BE}, \ref{ZERO_HBE},  \ref{ZERO_BEC} and  \ref{ZERO_HBEC}  and each product. The final two rows show the total $\text{lppd}^{z}$ across all products in the test and train sets, respectively.}
\label{table:zero_Gscores}
\centering	
\begin{tabular}{cccccccc}
\hline
Product $i$ & $\text{lppd}_{Base_{1},i}^{z,test}$ & $\text{lppd}_{HB,i}^{z,test}$ & $\text{lppd}_{BE,i}^{z,test}$& $\text{lppd}_{HBE,i}^{z,test}$ & $\text{lppd}_{BEC,i}^{z,test}$  & $\text{lppd}_{HBEC,i}^{z,test}$ \\
\hline
$1$ & -0.37 & -3.16 & -0.32 & -2.04& \bf{-0.32} & -1.97 \\ 
  $2$ & -73.47 & -65.66 & -60.85 & -55.87& -60.42 &  \bf{-55.18}  \\ 
  $3$ & -7.33 & -6.81 & -6.18 & \bf{-5.56} &  -6.23 & -5.59  \\ 
  $4$ & -29.44 &\bf{ -28.27} & -29.30 & -28.54 & -29.00 & -28.35  \\ 
  $5$ & -14.16 & -13.09 & -10.46 & -12.12 &\bf{-10.27} & -11.81\\ 
  $6$ & -3.67 & -5.80 & -2.55 & -3.63& \bf{-2.54} & -3.63  \\ 
  $7$ & -6.92 & -7.42 & \bf{-5.91} & -5.98  & -6.00 & -6.07 \\ 
  $8$ & -6.74 & -8.95 & -6.47 & -6.91 & \bf{-6.42} & -6.77 \\ 
  $9$ & -5.97 & -7.27 & \bf{-5.68} & -5.98 & -5.69 & -5.93  \\ 
  $10$ & \bf{-9.91} & -11.30 & -10.76 & -10.45 & -10.60 & -10.22  \\ 
  $11$ & -17.16 & -\bf{11.48} & -14.01 & -11.79 & -13.97 & -11.80\\ 
  $12$ & \bf{-9.80} & -11.86 & -10.48 & -10.53 & -10.30 & -10.27  \\ 
  $13$ & -15.84 & -15.25 & \bf{-9.75} & -9.99 & -9.81 & -9.91  \\ 
  $14$ & -10.34 & \bf{-8.66} & -11.15 & -9.93 & -11.11 & -9.95  \\ 
  $15$ & \bf{-10.36} & -11.15 & -10.78 & -10.49 & -10.83 & -10.52  \\ 
  $16$ & \bf{-5.61} & -7.47 & -6.12 & -6.60 & -6.19 & -6.61  \\ 
  $17$ & -15.01 & -15.23 & -13.60 & -13.09 & -13.66 & \bf{-13.07} \\  \hline
  $ \text{lppd}_{model}^{z, \text{test}}$ &  -242.10 & -238.82 & -214.37 & -209.50  & -213.35 & \bf{-207.65} \\  \hline \hline
     $ \text{lppd}_{model}^{z,\text{train}}$  &-708.26 & -699.89 & -609.45 & -662.65 & \bf{-608.48} & -662.84 \\ \hline
  \end{tabular}
  \centering
\end{table*}

\begin{figure}[H]
  \centering
     \includegraphics[trim={0 0.5cm 0 2.0cm},clip,width=0.49\textwidth]{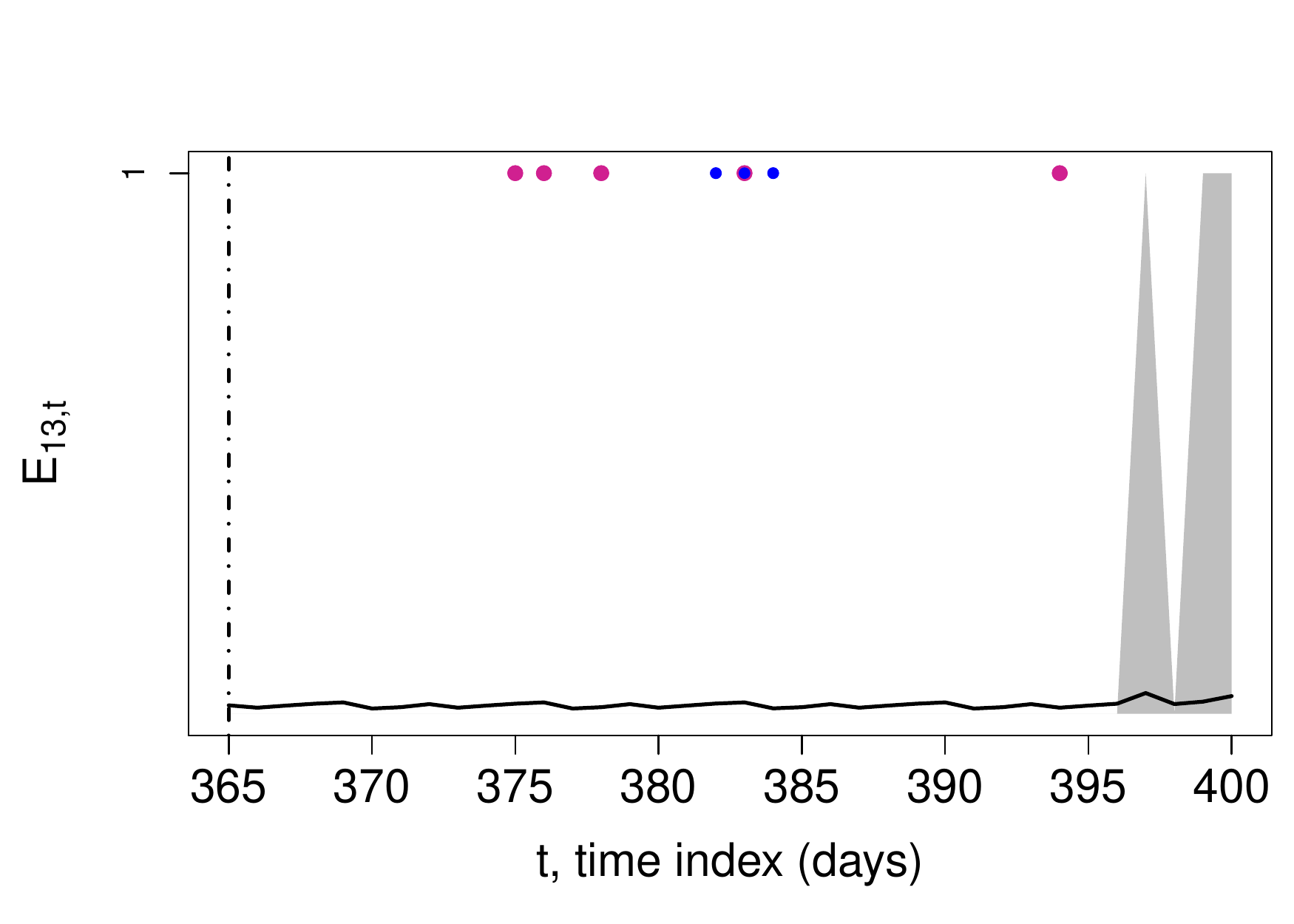}
     \includegraphics[trim={0 0.5cm 0 2.0cm},clip,width=0.49\textwidth]{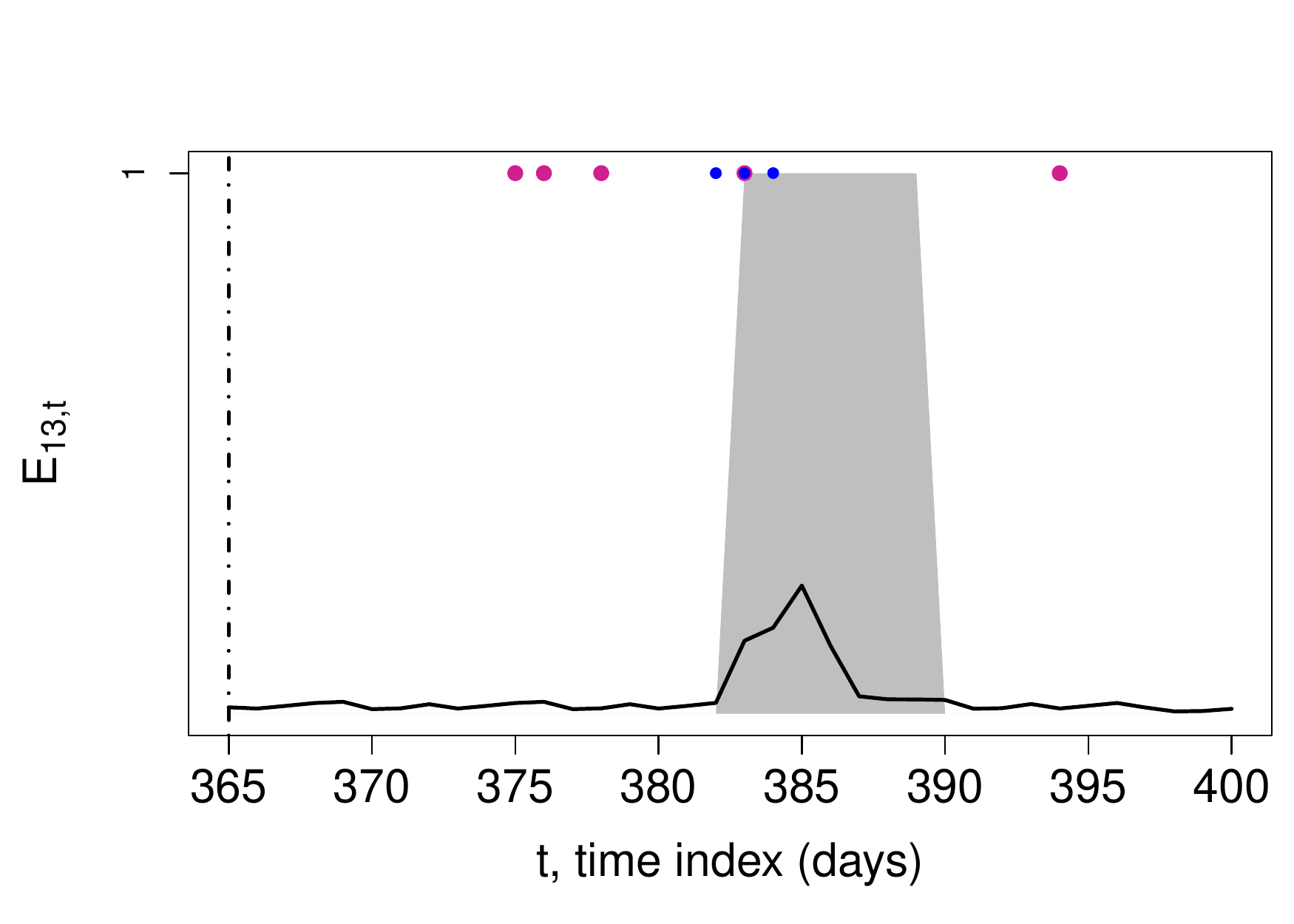} 
    \caption{\label{plot:ZERO_HBEvsZERO_HB} Plots of the predictive models \ref{ZERO_HB} (left) and \ref{ZERO_HBE} (right) for product $i=13$ over a portion of the test set. The blue and magenta dots represent self and cross event days respectively (i.e.~$E_{it}$ and  $\tilde{E}_{it}$). The black line is the estimated posterior mean of an event day observation (i.e.~$p_{it}$) and the shaded region is the $95\%$ credible interval of these estimates.}
\end{figure}

Table \ref{table:zero_Gscores} reveals some interesting findings. Firstly, we observe  the model \ref{ZERO_HB}, the zero process model with covariate information, provides a significant  improvement in predictive performance compared to baseline model \ref{ZERO_NM1} without covariate information. We further see that inclusion of a self-excitation component in (\ref{ZI_model}) provides a marked improvement over the model \ref{ZERO_HB} without self-excitation.  Figure \ref{plot:ZERO_HBEvsZERO_HB} demonstrates an example of the benefit of self-excitation inclusion by comparing the event day prediction performance between models \ref{ZERO_HBE} and \ref{ZERO_HB} over a portion of the test set. We observe inclusion of self-excitation produces a $95\%$ credibility interval of model \ref{ZERO_HBE}  that captures a subsequent sale that model \ref{ZERO_HB}  does not immediately after the first sale at $t=382$.

\begin{figure}[H]
  \centering
     \includegraphics[trim={0 0.5cm 0 2.0cm},clip,width=0.49\textwidth]{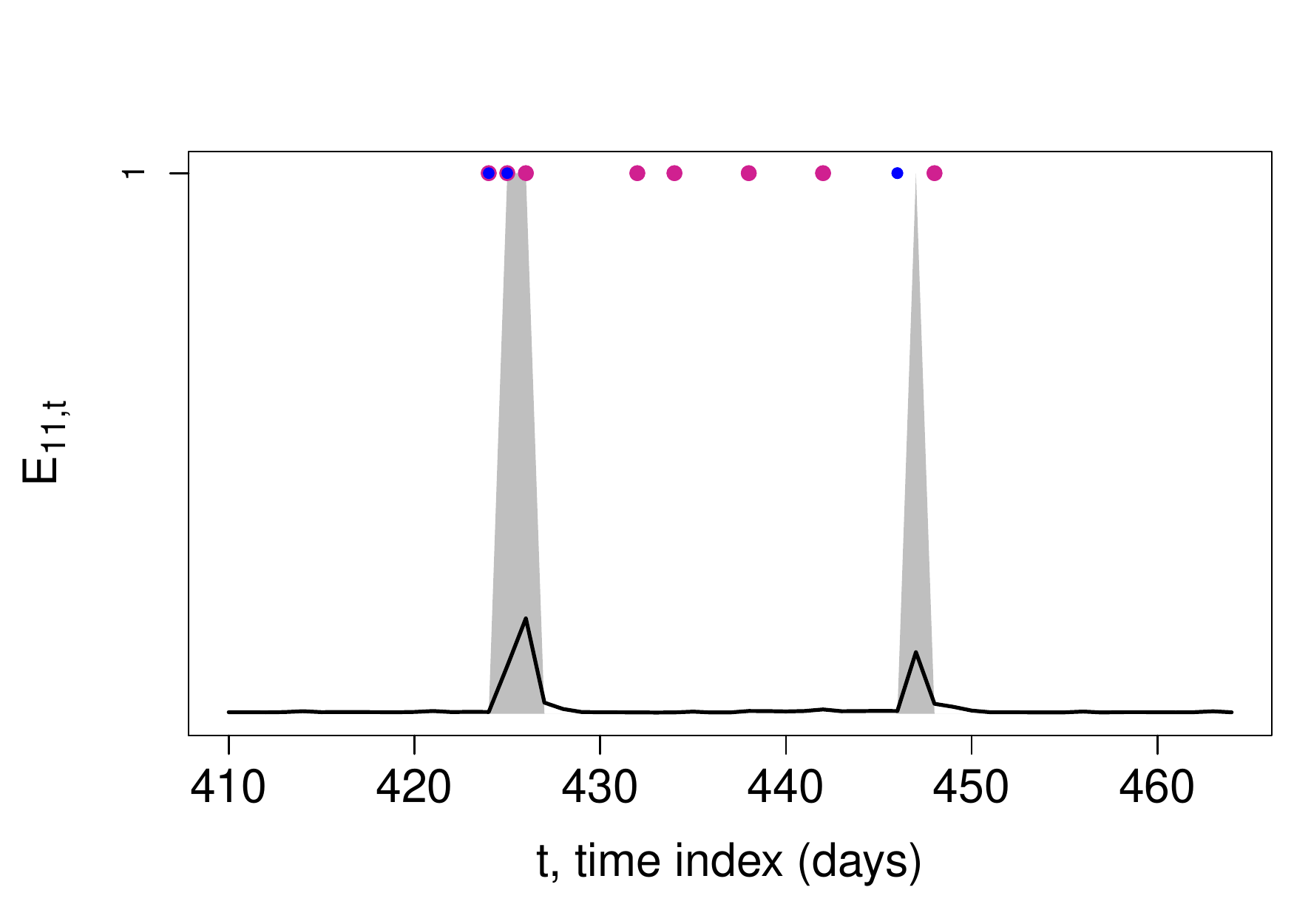}
     \includegraphics[trim={0 0.5cm 0 2.0cm},clip,width=0.49\textwidth]{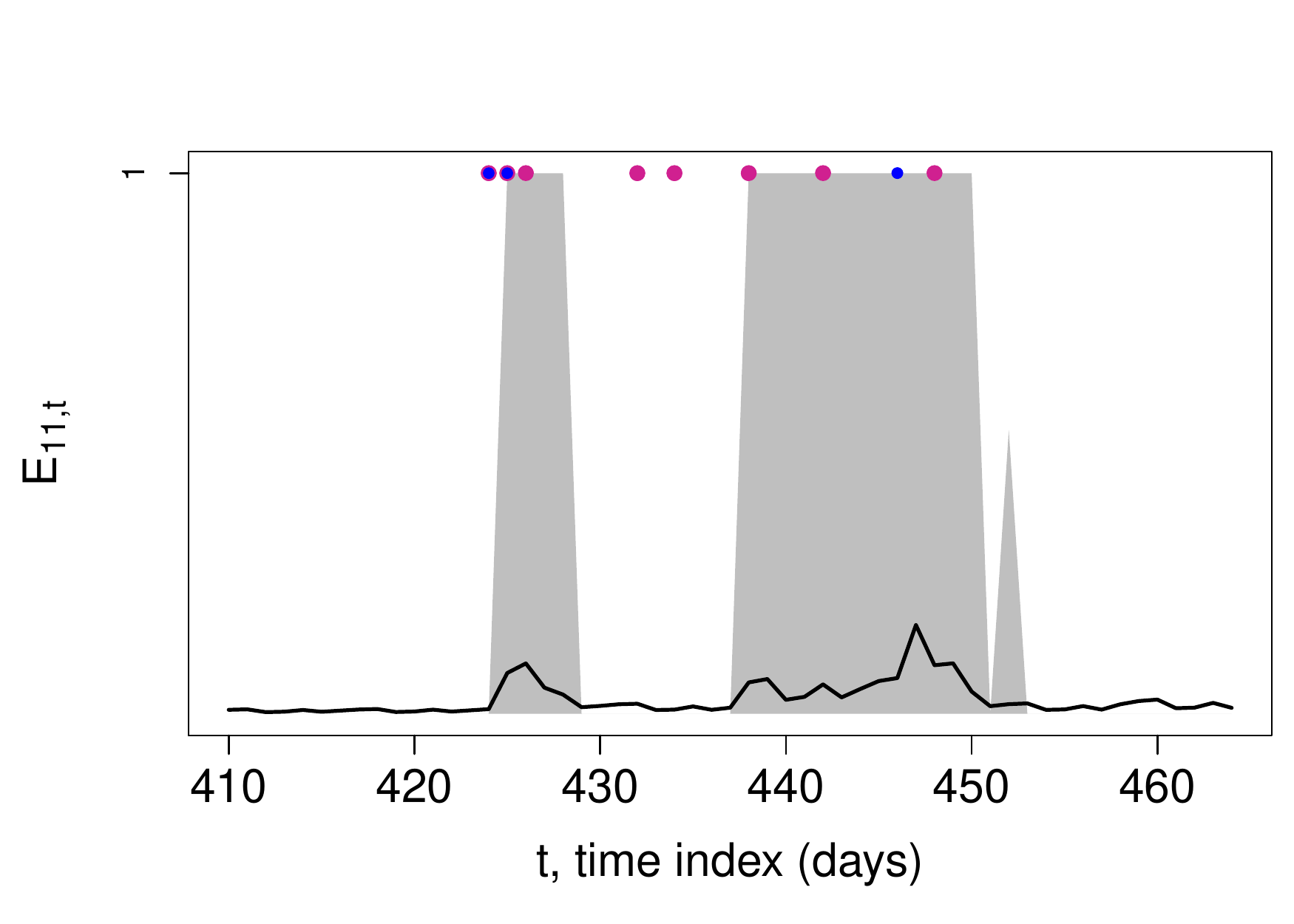} 
    \caption{\label{plot:ZERO_HBECvsZERO_HBE}Plots of the predictive models \ref{ZERO_BE} (left) and \ref{ZERO_HBE} (right) for product $i=11$ over a portion of the test set. The blue and magenta dots represent self and cross event days respectively (i.e.~$E_{it}$ and  $\tilde{E}_{it}$). The black line is the estimated posterior mean of an event day observation (i.e.~$p_{it}$) and the shaded region is the $95\%$ credible interval of these estimates. }
\end{figure}

Table \ref{table:zero_Gscores} further indicates the predictive benefits that hierarchical extensions provide over its non-hierarchical equivalents. Figure \ref{plot:ZERO_HBECvsZERO_HBE} illustrates an example of the benefit of these hierarchical extensions by comparing event day prediction performance between models \ref{ZERO_HBE} and \ref{ZERO_BE} over a portion of the test set. We observe that  by information pooling across the intermittent demand series produces a $95\%$ credibility interval of model \ref{ZERO_HBE} that  captures a sale at $t=446$ (during the Christmas period). This is despite the absence of sales over the Christmas period of the previous year for this product. In this way, the hierarchical model benefits from inferring parameter values of other intermittent demand series which have observed sales over the previous the Christmas period. 

Finally, Table \ref{table:zero_Gscores} indicates that the cross-excitation expositions of models  \ref{ZERO_BEC} and \ref{ZERO_HBEC}  offer an improvement in event day prediction over the test set compared to their non cross-excitation counterparts (i.e.~\ref{ZERO_BE} and \ref{ZERO_HBE}). Interestingly, cross-excitation does not offer any benefits in terms of the training set; but shows significant predictive gains in the test set.

\subsection{Count process variations}\label{Count_process_variations}
Similarly to section \ref{zero_fits}, the benefits of the excitation and hierarchical component to the count process of hurdle model (\ref{ZI_model}) are verified by implementing the following cumulative variations in the link functions and hierarchical layerings of the model for each  $i=1,\ldots, 17$. These model variations follow the same rationale  as with the zero process.

\begin{itemize}
\item \otherlabel{COUNT_NM1}{Base$_{1}^{c}$}  \bf  Baseline model (Base$_{1}^{c}$)\rm: We learn the count process  with link function $$  \log(\lambda(\boldsymbol{x}^c_{it},H_{it},  \boldsymbol{\theta}^c_{i}) ) =  \varphi_{i}^{c},$$  
i.e.~a constant rate per product.  This is the Bayesian baseline model as it estimates the zero-process independent of covariate information. The $\varphi_{i}^{c}$ is estimated using vague priors. 

\item \otherlabel{COUNT_HB}{HB$^{c}$}  \bf  Hierarchical Bayesian  (HB$^{c}$)\rm: We learn the count process  with link function $$  \log(\lambda(\boldsymbol{x}^c_{it},H_{it},  \boldsymbol{\theta}^c_{i}) ) =  \varphi_{i}^{c}(t),$$  
with the hierarchical prior formulation discussed in section \ref{self_cross_hurdle_model}. 

\item \otherlabel{COUNT_BE}{BE$^{c}$}  \bf  Bayesian with self-excitation  (BE$^{c}$)\rm: We learn the count process   with  link function
$$  \log(\lambda(\boldsymbol{x}^c_{it},H_{it},  \boldsymbol{\theta}^c_{i}) ) =  \varphi_{i}^{c}(t)+S_{it}^{c},$$  
but exclude the hierarchical prior formulation shown in section \ref{self_cross_hurdle_model}. 

\item \otherlabel{COUNT_HBE}{HBE$^{c}$}  \bf  Hierarchical Bayesian with self-excitation  (HBE$^{c}$)\rm: This is the full model discussed in the section \ref{self_cross_hurdle_model}. We learn the count process with link function 
$$  \log(\lambda(\boldsymbol{x}^c_{it},H_{it},  \boldsymbol{\theta}^c_{i}) ) =  \varphi_{i}^{c}(t)+S_{it}^{c},$$  
with the hierarchical prior formulation discussed in section \ref{self_cross_hurdle_model}.
\end{itemize}

Parameter inference is performed by Hamiltonian Monte Carlo sampling algorithm and is implemented using the \textsf{rstan} library \citep{RSTAN}.  Convergence was confirmed by Heidelberger Welch statistic  across all models and parameters \citep*{heidelberger1981spectral}. The specification of these hyper-priors and constant of models  \ref{COUNT_HB}, \ref{COUNT_BE} and \ref{COUNT_HBE}  is included in appendix \ref{priors}.  For further MCMC implementation details, as well as additional model comparisons and discussion, refer to the supplementary material.

\subsection{Count process fits}\label{count_fits}
Similarly to the zero processes outlined in section \ref{zero_fits}, we test the performance of the count variation models \ref{COUNT_NM1}, \ref{COUNT_HB}, \ref{COUNT_BE} and \ref{COUNT_HBE} by calculating how capable each model is of predicting the volume of sales on event days (i.e.~days when sale has been observed) over the test interval $T^{test}$ (between $30^{th}$ September 2014 to $7^{th}$ January 2015) for each $i=1,\ldots, 17$ given the history of self events $H_{it}$, covariate information $ \boldsymbol{x}^c_{it}$ and posterior samples. We apply the same methodology over the test interval as with the zero process:

\begin{enumerate}

\item On event day $t$ (i.e.~$E_{t}=1$) on the test interval and $s^{th}$ posterior sample, we compute the full predictive posterior distribution of the volume of sales occurring conditioned on $ H_{it}, \boldsymbol{x}^c_{it} ,  \boldsymbol{\theta}^{c}_{is}$ for each $i=1,\ldots, 17$.

\item We observe $y_{it+1}$ (the volume of sales of product $i$ on day $t+1$) for each $i=1,\ldots, 17$ and update the self event histories $H_{it+1}$ for $i=1,\ldots, 17$.

\item Repeat steps for each $t $, for each sample $s$ and $i$ over the test period of $30^{th}$ September 2014 to $7^{th}$ January 2015.

\end{enumerate}
This builds up a set of posterior rates $\lambda_{its}$ for samples  $s=1,\dots,S $ for the probability of the number of sales on a given event day over $T^{test}$ for each $i=1,\ldots, 17$ based on our posterior sample fits inferred from $T^{train}$ conditioned on $ \boldsymbol{x}^c_{it} ,H_{it},   \boldsymbol{\theta}^c_i$.
\newline

Similarly to the zero process, we evaluate the predictive performance by calculating the log posterior predictive density for each of the products $i=1,\ldots, 17$. The  log posterior predictive density $\text{lppd}^{c}$ for the count process is given by:
 $$\text{lppd}_i^{c} =  \sum_{t \in T_{i}} \log\left( \frac{1}{S}\sum_{s=1}^{S}\binom {y_{ik}-2+\phi}{y_{ik}-1} \left(\frac{\lambda_{its}-1} {\lambda_{its}-1+\phi}\right)^{y_{ik}-1}\left(\frac{\phi} {\lambda_{its}-1+\phi}\right)^{\phi}  \right)    $$
where $\phi=1$ and $\lambda_{its}$  is the prediction mean of count sales occurring for product $i$ from the $s^{th}$ posterior sample  for some model of interest and  $T_i = \{ t |  y_{it}>0  \}$, i.e.~$T_i$ are the set time indices corresponding to sales days for product $i$ over some interval of time. Table \ref{table:count_Gscores} provides the $\text{lppd}^{c} $ scores for across products and models.

\begin{table*}
\caption{$\text{lppd}^{c}_{i}$ scores of the count process fits for the models \ref{COUNT_NM1}, \ref{COUNT_HB},   \ref{COUNT_BE} and \ref{COUNT_HBE} for each product and fitted model. The final two rows show the total $\text{lppd}^c$ across all products in the test and train sets, respectively.}
\label{table:count_Gscores}
\begin{tabular}{ccccc}
\hline
Product $i$ & $\text{lppd}_{Base_{0},i}^{c,test}$ & $\text{lppd}_{HB,i}^{c,test}$ & $\text{lppd}_{BE,i}^{c,test}$  & $\text{lppd}_{HBE,i}^{c,test}$  \\
\hline
$1$ & 0.00 & 0.00 & 0.00 & 0.00  \\ 
  $2$ & -18.10 & -18.78 & \bf{-13.59} & -14.18 \\ 
  $3$ & -0.91 &  -0.55 & -0.62 & \bf{-0.48}  \\ 
  $4$ & \bf{-1.60} & -1.78 & -1.66 & -1.77  \\ 
  $5$ & -0.08 & \bf{-0.07} & -0.08 & -0.66\\ 
  $6$ & 0.00 & 0.00 & 0.00 & 0.00 \\ 
  $7$ & -0.01 & \bf{-0.00} & -0.04 & -0.22 \\ 
  $8$ & -4.99 & -4.16 & -7.92 & \bf{-3.17}  \\ 
  $9$ & -2.54 & -\bf{1.40} & -1.50 & -1.60\\
  $10$ & -3.98 & -3.98 & -3.80 & \bf{-2.04}  \\ 
  $11$ & \bf{-7.05} & -7.07 & -7.45 & -10.95 \\ 
  $12$ & -1.02 & -1.09 & -1.03 & \bf{-0.68}  \\ 
  $13$ & -3.46 & -3.47 & -3.47 & \bf{-2.33}\\ 
  $14$ & -6.19 & 6.46 & -6.48 &  \bf{-5.23} \\ 
  $15$ & -2.04 & -2.05 & -1.95 & \bf{-0.66}\\ 
  $16$ & \bf{-1.57} & -2.64 & -1.63 & -1.80\\ 
  $17$ & -0.10 & \bf{-0.08} & -0.09& -0.55 \\ 
      \hline
  $ \text{lppd}_{model}^{c, \text{test}}$ & -53.64 & -53.60 & -51.32 & \bf{-46.33}  \\ \hline\hline
  $\text{lppd}_{model}^{c, \text{train}}$ &-336.81 & -335.21 & \bf{-308.58} & -325.15\\ \hline
\end{tabular}
\end{table*}

 Table \ref{table:count_Gscores} reveals some interesting findings. Firstly, we observe that the model variations of  \ref{COUNT_HB}, \ref{COUNT_BE} and \ref{COUNT_HBE}  perform significantly better than the Baseline model  \ref{COUNT_NM1} with no covariates.  Similarly  to the zero process, Table \ref{table:count_Gscores} indicates the count process uniformly benefits from the inclusion of self-excitation in the model variations outlined in \ref{Count_process_variations}.

\begin{figure}[t]
  \centering
     \includegraphics[trim={0 0.5cm 0 2.0cm},clip,width=0.49\textwidth]{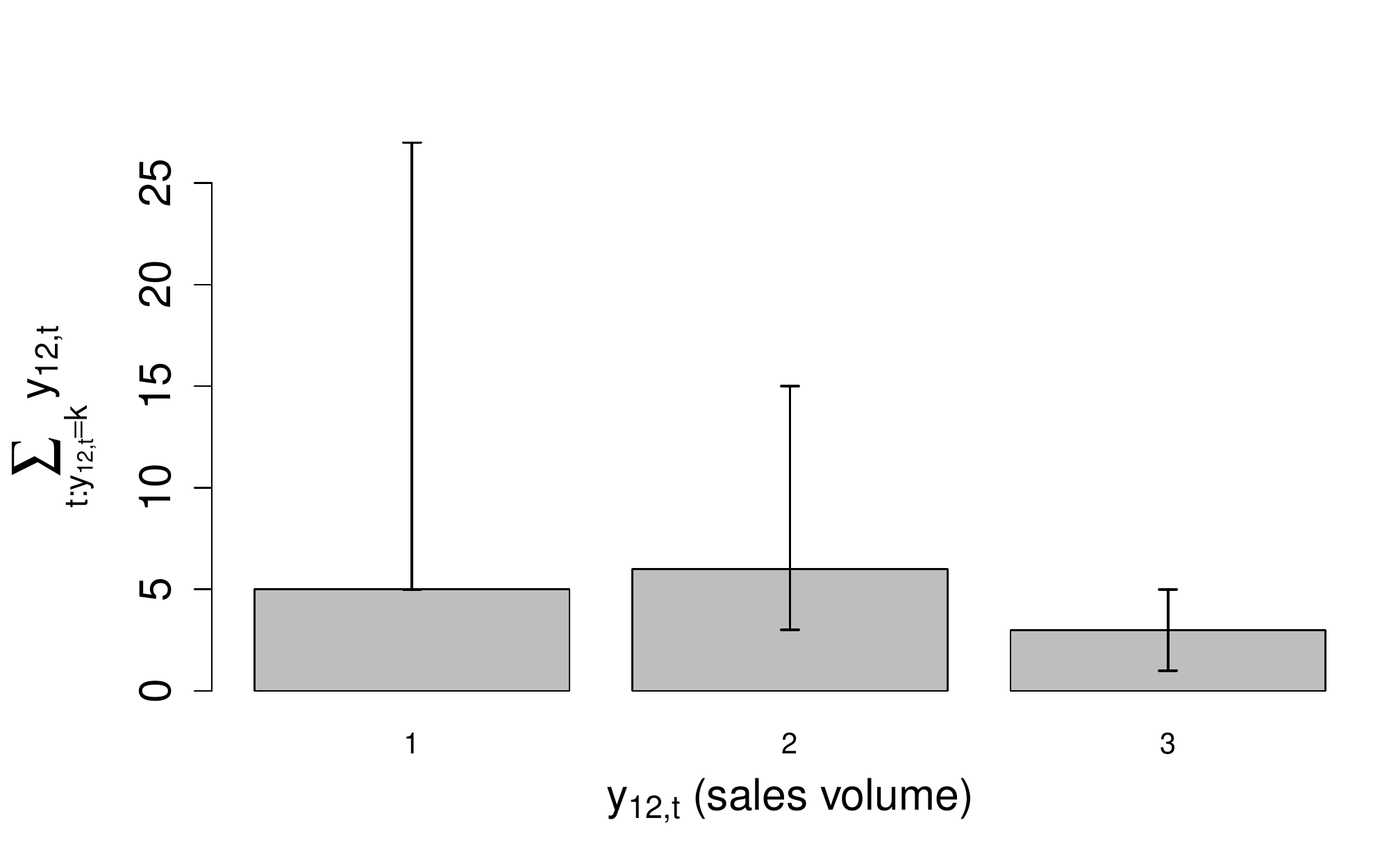}
     \includegraphics[trim={0 0.5cm 0 2.0cm},clip,width=0.49\textwidth]{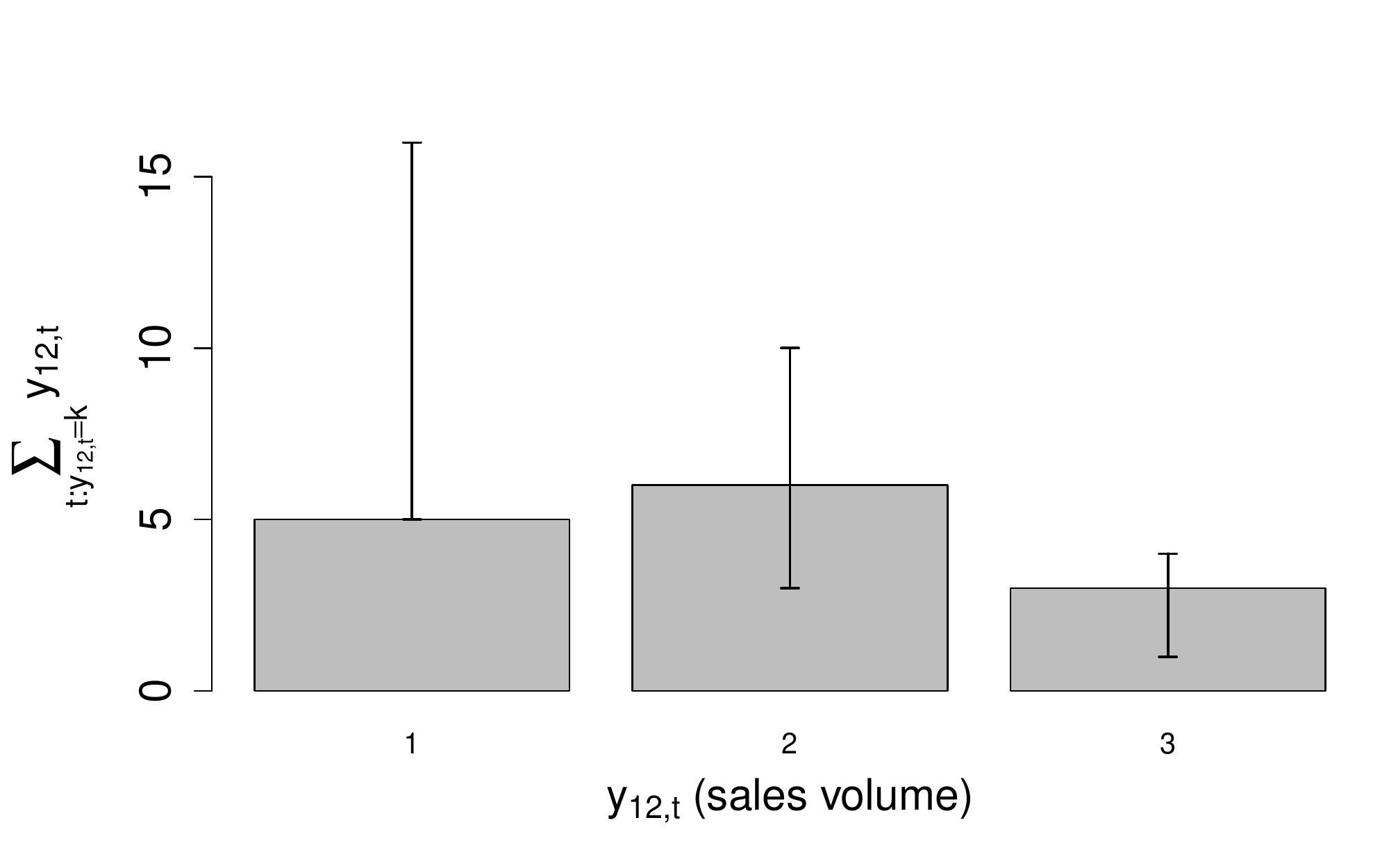} 
    \caption{\label{plot:COUNT_HBvsCOUNT_BEC}  Histograms  of observed $\sum_{t:y_{it}=k}y_{it}$  with corresponding 95\% credible intervals of the posterior predictive distributions for models \ref{COUNT_BE} (left) and \ref{COUNT_HBE} (right) for product $i=12$.  The lower of $2.5\%$ credible interval (the lower bound of the whisker bars) for $\sum_{t:y_{it}=1}\tilde{y}_{it}$ will at best be $\sum_{t:y_{it}=1}1$, since the count distribution is lower bounded by 1.}
\end{figure}

 We further see that the count process benefits more from the hierarchical borrowing across the intermittent demand series. This is understandable given the level of sparsity in the count process. As Table \ref{table:sale_rates} indicates, the order of sales that the each intermittent demand series has is very small (typically in the order 3-20 sales), and thus it may be expected that information borrowing would particularly benefit the individual models. An example of this additive strength of the hierarchical exposition of the count model variations is illustrated by Figure \ref{plot:COUNT_HBvsCOUNT_BEC}. This plot shows a histogram of $y_{it}$ against  the sum of $\sum_{t:y_{it}=k}y_{it}$ (for product 12) with corresponding 95\% credibility intervals of posterior predictive distributions for the models \ref{COUNT_HB} and \ref{COUNT_BE}. We observe that the hierarchical model variation (even without the excitation) produces much tighter credibility intervals around the observed data than the model without information borrowing.

\begin{figure}[t]
  \centering
     \includegraphics[trim={0 0 0 2.0cm},clip,width=0.49\textwidth]{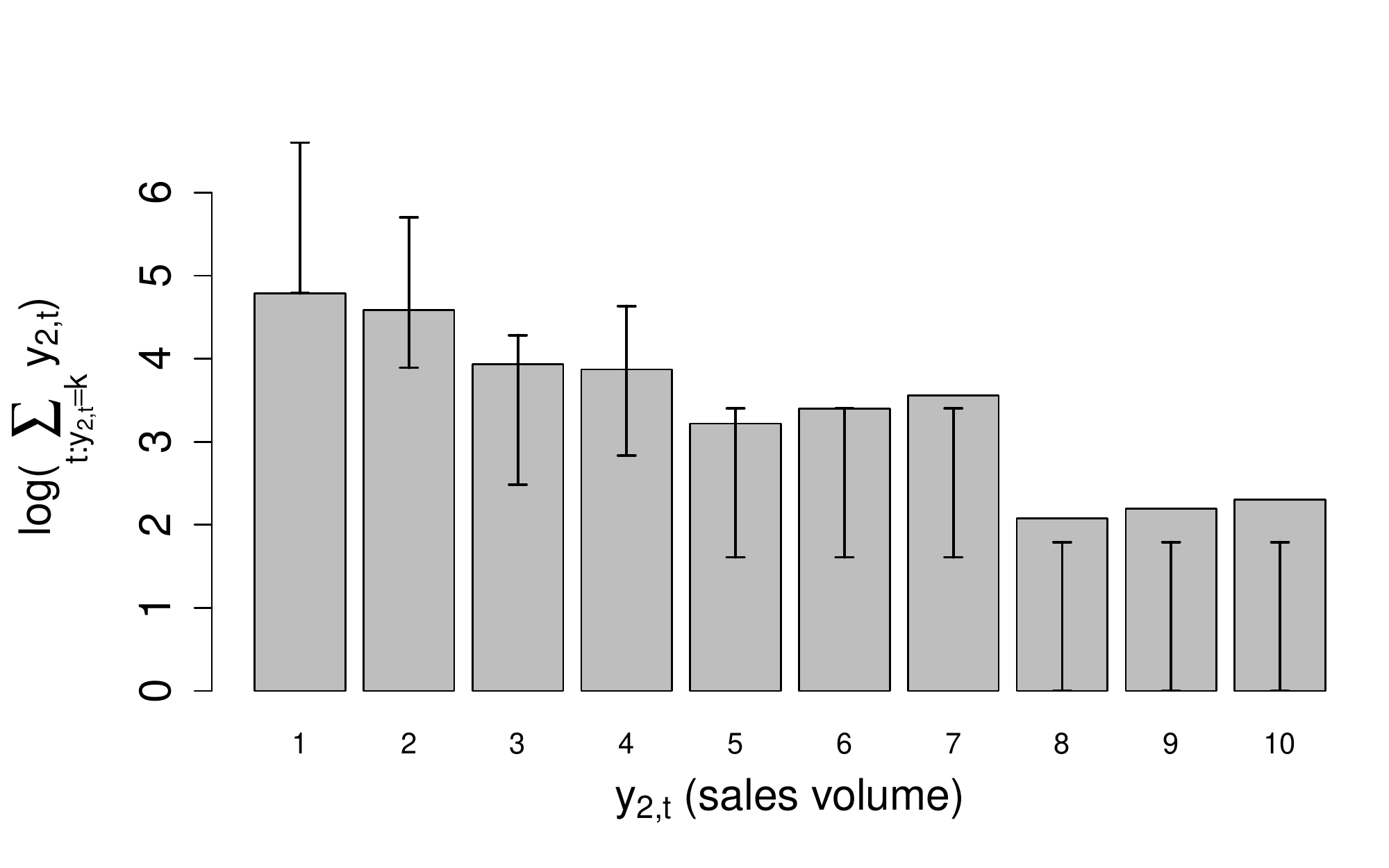}
     \includegraphics[trim={0 0 0 2.0cm},clip,width=0.49\textwidth]{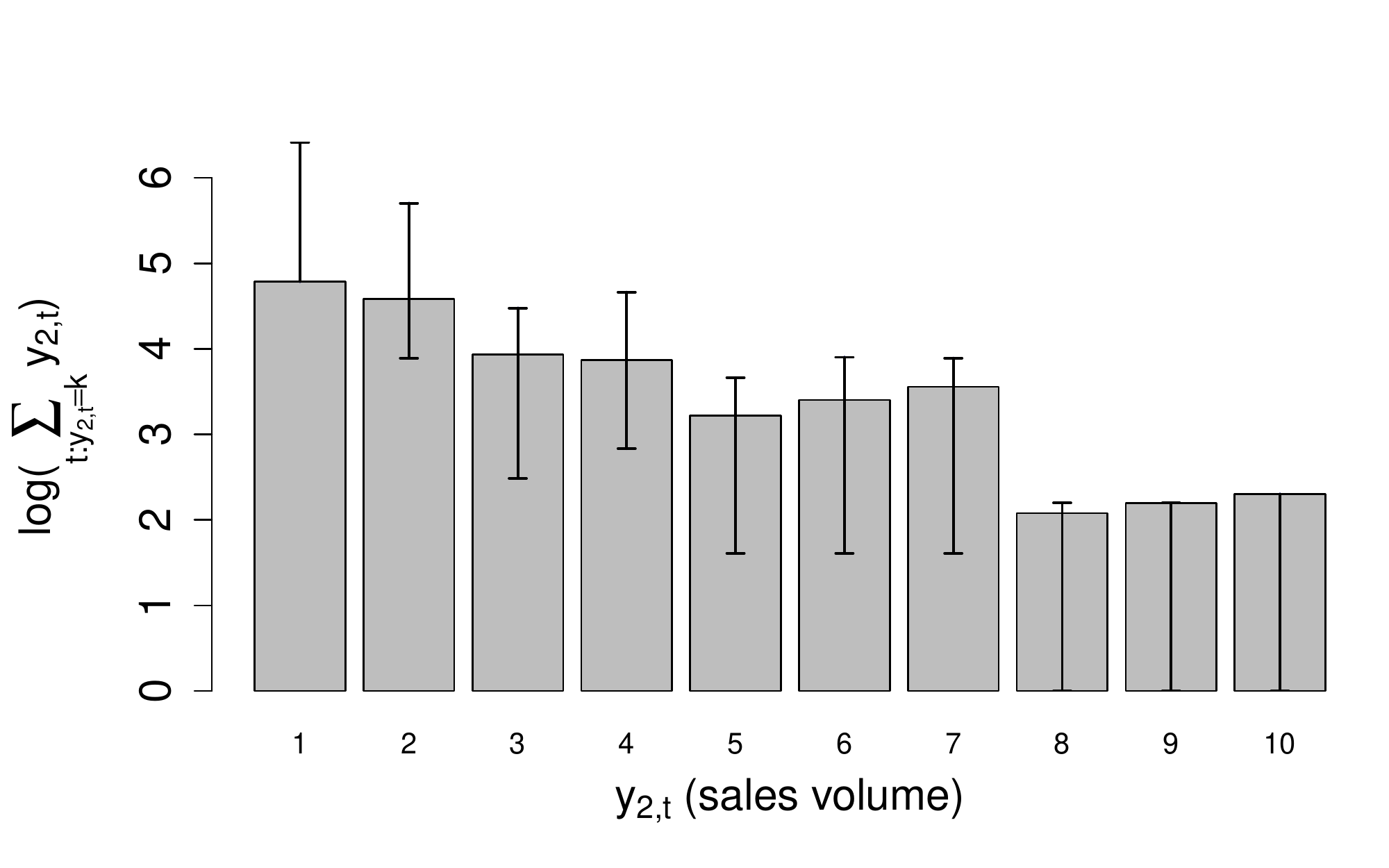} 
    \caption{\label{plot:COUNT_HBvsCOUNT_HBE}   Histograms  of observed $\sum_{t:y_{it}=k}y_{it}$  with corresponding 95\% credible intervals of the log of the posterior predictive distributions of $\sum_{t:y_{it}=k}\tilde{y}_{it}$ (sale counts) for models \ref{COUNT_HBE} (left) and \ref{COUNT_HB} (right) for product $i=2$.  }
\end{figure}

However, the best performing models are ones with both information borrowing and self-excitation. Figure \ref{plot:COUNT_HBvsCOUNT_HBE} illustrates the optimal performance of  \ref{COUNT_HBE} over \ref{COUNT_HB}. In this plot, we see  the 95\% credibility intervals produced from model  \ref{COUNT_HBE}  for the higher count instances ($7+$) capture the observed aggregated count instances, whereas the \ref{COUNT_HB} credibility intervals fail to do so. We further see the aggregate log posterior predictive density of $ \sum_{i=1}^{17} \text{lppd}_{model,i}^{c, \text{train}}$ of Table \ref{table:count_Gscores} provides more evidence that model \ref{COUNT_HBE} is the best fitting model.

\subsection{Retail analytics discussion}\label{SMG:retailanalytics_discussion}
The output of models outlined in sections \ref{Zero_process_variations} and \ref{Count_process_variations} provides interesting interpretations from a retail analytics perspective. Firstly, we observe that covariate data $\boldsymbol{x}^z_{it}, \boldsymbol{x}_{it}$ as specified in \ref{SMI:covariates} improves forecasting performance for the intermittent demand series of SMI products. This is  indicated in both  \ref{COUNT_HB} and \ref{ZERO_HB} - models with regression parameters and no form of excitation - outperforming their baseline counterparts on both the training and test sets. This importantly sheds light into the intermittent demand of SMI, in that it demonstrates covariate data such as prices and seasonality ought to be incorporated into training forecasting models as it seems predictions are improved from their inclusion.

\begin{figure}[t]
   \centering
       \includegraphics[trim={0 0.5cm 0 2.0cm}, clip, width=0.9\textwidth]{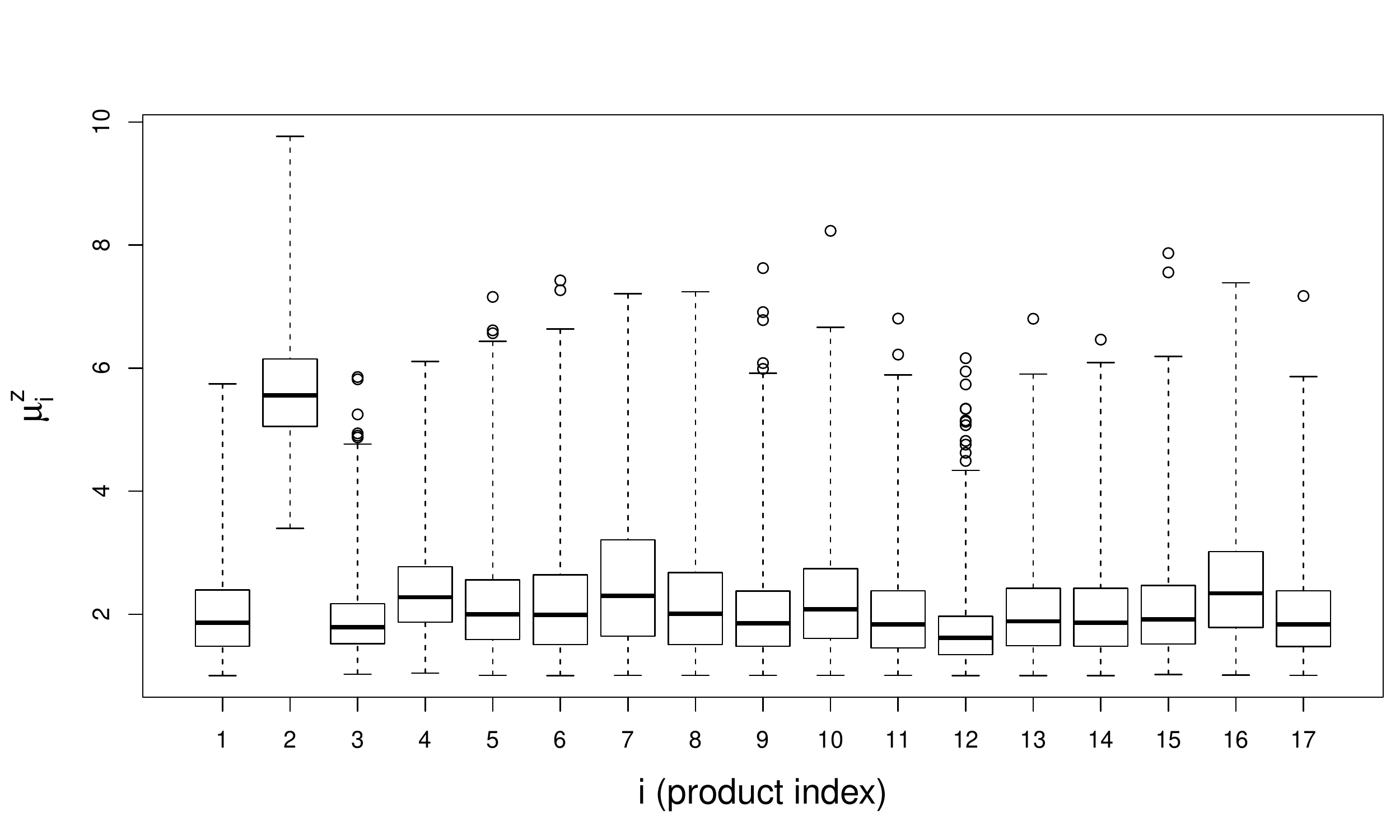}
 
         \caption{Box plots of the posterior distribution of $\mu_{i}^c$  across all products for model \ref{ZERO_HBE}. The $\mu_{i}^c$ estimates being greater than 2 indicates the temporal excitation exhibited in that data  typically occurs at lags greater than 1.  }\label{plot:ZEROboxplots2}
 \end{figure}

Our findings further support the hypothesis that intermittent demand forecasting is improved when excitation dynamics are incorporated into models. This supports the findings of \cite{snyder2012forecasting} and \cite{chapados2014effective} where they establish that models incorporating  the recent sales history outperform temporally static models. This is important because it ultimately allows retailers to circumvent over-stocking that typically results from inaccurate forecasting \citep*{ghobbar2003evaluation}. However, our findings reveal some aspects of intermittent demand forecasting that goes beyond the work of \cite{snyder2012forecasting} and  \cite{chapados2014effective}. Namely, we establish that the temporal excitation exists even if you condition on the seasonal trends and pricing information of $\boldsymbol{x}_{it}$. This suggests that temporal excitation is systematic and occurs beyond the variables traditionally utilised in forecasting models. We furthermore find that temporal excitation is manifested at lags greater than 1. Figure \ref{plot:ZEROboxplots2} demonstrates that $ \mu^{z}_{i}$ (the mean of excitation function of $g(\cdot \mid  \mu, \tau)$) is approximately 2 across the majority of products, which implies that $2/3$ of the probability mass of $g(\cdot \mid  \mu, \tau)$ is placed on lags greater than or equal to $2$. This is crucially important, as it indicates that a simple $AR(1)$ (or similar) is possibly not enough compared to the Hawkes process that incorporates the entire history of events.

\begin{figure}[t]
  \centering
     \includegraphics[trim={0 0.5cm 0 2.0cm},clip,width=0.49\textwidth]{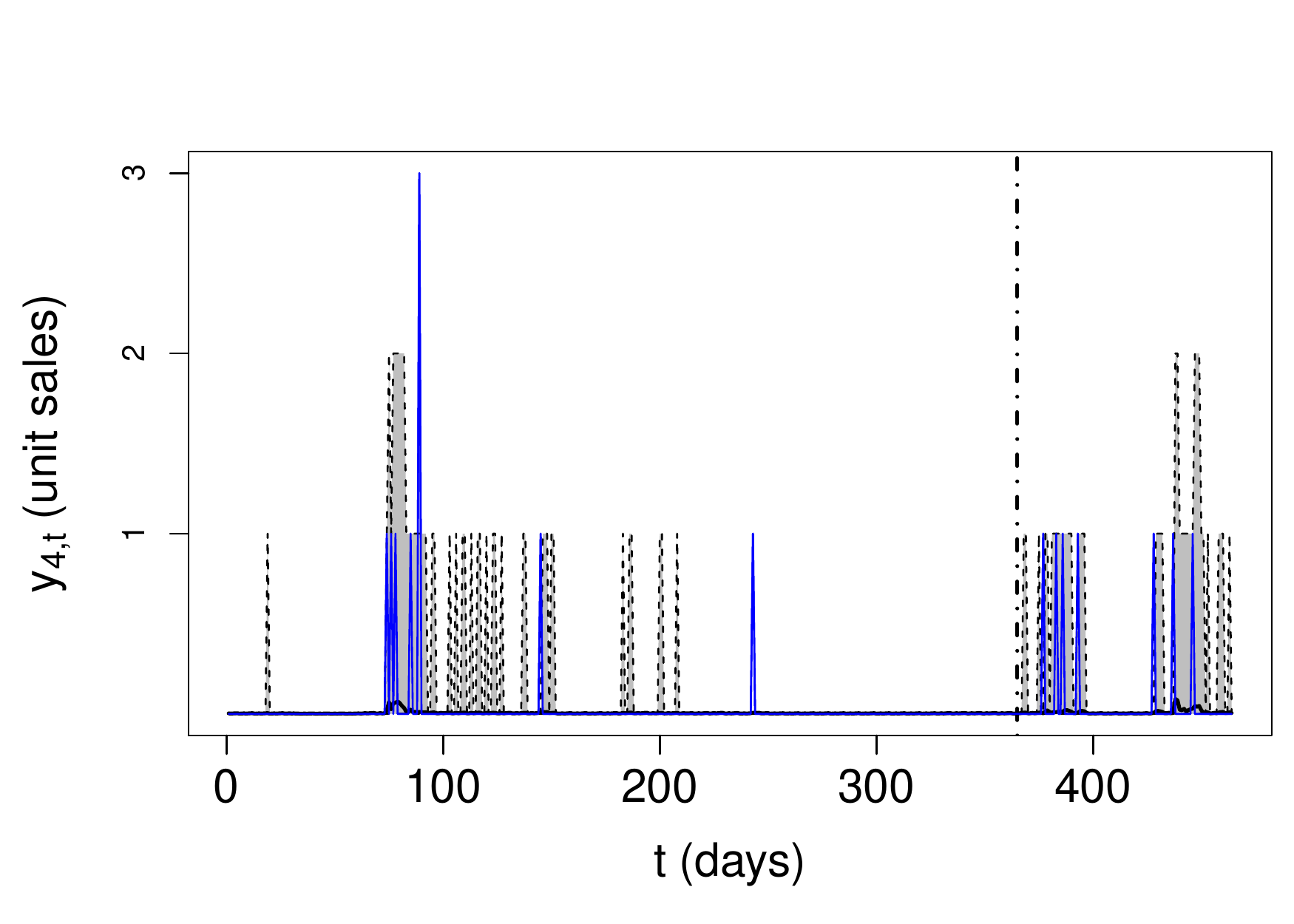}
     \includegraphics[trim={0 0.5cm 0 2.0cm},clip,width=0.49\textwidth]{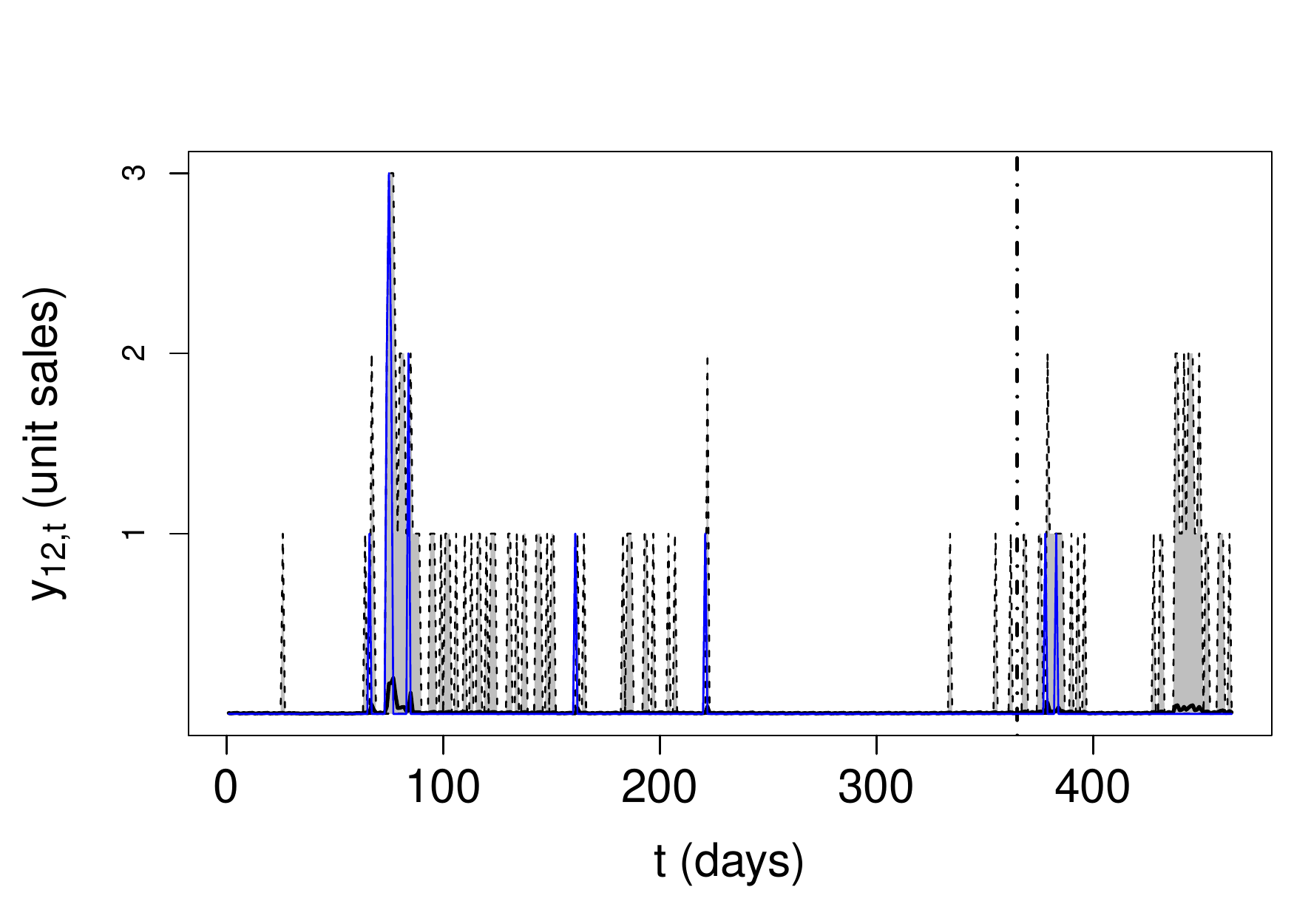} 
    \caption{Plots of the combined models \ref{ZERO_HBE} and \ref{COUNT_HBE}  for product $i=4$ (left) and product $i=12$ (right) over the entire training and test sets. The blue solid lines correspond to the observed time series, whereas the shaded region and corresponding dashed black lines to 95\% prediction intervals. The vertical line at $t=365$ represents the end and start of the training and test sets respectively.  \label{Combine_zeroANDcountpred}}
\end{figure}

Finally, Figure \ref{Combine_zeroANDcountpred} shows the sales forecasts of two slow-moving-inventory products using the combined zero and count models. Despite the severe lack of data within each of the time series, our model is able to produce meaningful predictions in the test set, including prediction intervals, capturing several of the observed sales.

\section{Conclusion}\label{SMG:Conclusion}
In this work we introduced a hierarchical model for the sales of the slow-moving-inventory category of touchscreen tablets across five large supermarkets in south London. We modelled the sales process as a Bayesian hierarchical zero-inflated hurdle regression model with self and cross-excitation components. Our model specification is interpretable and allows a deeper understanding of the role that covariates, self-excitation and cross-excitation play in the sales process of slow-moving-inventory and further provides a fully specified predictive distribution over this process. We demonstrated that the hierarchical structure as well as the self and cross-excitation additions offer a significant improvement in the predictive accuracy of this SMI sales process.

This model has important implications to the challenging issues that retail analytics face when developing SMI models. Firstly, it offers utility in terms of demand and profit forecasting that will allow retailers more accurate predictions of the sales distributions to aid with the issue of inventory management as well as price optimisation over short term horizons. It helps to explain the sources of variation and uncertainty that is exhibited in  intermittent  demand processes that previously was not well understood. The model also reveals a strong excitation component to these sales which could warrant further investigation into potential underlying factors that could explain the observed excitation  (e.g. marketing campaigns). We further note that, though there are many other approaches of specifying the cross-excitation relationship between pairwise products, our adopted approach of cross-excitation within brand provides an intuitive and computationally simple method of expressing the suspected temporal cross-correlation.

This work could be extended in many different directions. For example, a variable selection methodology could be introduced into the covariate predictors for each of the regression models. Our approach specified a priori the cross-excitation structure by defining an excitation event as an  a sale occurring within the same brand; it could also be interesting to assess whether the excitation structure could  be inferred from the data.

\clearpage
\appendix

\section{Appendix}\label{app}

\subsection{\label{priors}Prior formulation}
Table \ref{table:ZERO_prior_formulations} specifies the prior structure of the zero process models models \ref{ZERO_NM1}, \ref{ZERO_HB},  \ref{ZERO_BE},  \ref{ZERO_HBE},   \ref{ZERO_BEC},  and \ref{ZERO_HBEC}.  Table \ref{table:COUNT_prior_formulations} specifies the prior structure of the count process models  \ref{COUNT_NM1},  \ref{COUNT_HB}, \ref{COUNT_BE} and \ref{COUNT_HBE}.

\begin{table*}[h]\label{zero_prior_table}
\caption{Prior formulation of models \ref{ZERO_NM1}, \ref{ZERO_HB},  \ref{ZERO_BE},  \ref{ZERO_HBE},   \ref{ZERO_BEC} and  \ref{ZERO_HBEC}.  We abbreviate Normal$(\mu^2,\sigma)$ and Gamma$(\alpha, \beta  )$ to N$(\mu,\sigma^2)$ and G$(\alpha, \beta  )$ respectively.}
\label{table:ZERO_prior_formulations}
\hspace*{-1.2cm}
\begin{tabular}{ccccccc}
\hline
Parameter & \ref{ZERO_NM1} & \ref{ZERO_HB} &    \ref{ZERO_BE} &   \ref{ZERO_HBE} &    \ref{ZERO_BEC}  &  \ref{ZERO_HBEC} \\
\hline
 $\varphi_{i} \sim $  & $\text{N}(-3, 3) $ & &  &  & &    \\ 
 $\theta^z_{i1} \sim $  & & $\text{N}(\mu_{1}^{z}, 0.05) $ & $\text{N}(-3,0.75) $ & $\text{N}(\mu_{1}^{z}, 0.05)$  &  $\text{N}(-3,0.75) $ & $\text{N}(\mu_{1}^{z}, 0.05)$    \\ 
 $\theta^z_{i2} \sim $  & & $\text{N}(\mu_{2}^{z}, 0.05) $ & $\text{N}(0, 0.75) $& $\text{N}(\mu_{2}^{z}, 0.05)$  &  $\text{N}(0, 0.75) $ & $\text{N}(\mu_{2}^{z}, 0.05)$    \\ 
   \vdots &    &\vdots &  \vdots &  \vdots &     \vdots &  \vdots    \\ 
 $ \theta^z_{i20} \sim$  & & $\text{N}(\mu_{20}^{z}, 0.05)$  & $\text{N}(0, 0.75) $& $\text{N}(\mu_{20}^{z}, 0.05)$ &  $\text{N}(0, 0.75)$  & $\text{N}(\mu_{20}^{z}, 0.05)$     \\[0.3cm]   
  
   $ \gamma_{i1}^{z} \sim $  & &   & $\text{G}(5, 1)$& $\text{G}(\eta_{1}^{z}, 1)$ &  $\text{G}(5, 1)$ & $ \text{G}(\eta_{1}^{z}, 1)$ \\ 
    $\gamma_{i2}^{z} \sim$  & &  &$1+\text{G}(1, 2)$ & $1+\text{G}(\eta_{2}^{z}, 2)$  &$1+\text{G}(1, 2)$ & $1+\text{G}(\eta_{2}^{z}, 2)$   \\ 
   $\gamma_{i3}^{z} \sim$  & &  &  $\text{G}(10, 2.5)$ &$\text{G}(\eta_{3}^{z}, 2.5)$  & $\text{G}(10, 2.5)$ & $\text{G}(\eta_{3}^{z}, 2.5)$   \\[0.3cm]   
    $ \widetilde{\gamma}_{i1}^{z}  \sim$  & &   &    && $\text{G}(2, 8)$ & $\text{G}(\widetilde{\eta}_{1}^{z}, 8)$  \\ 
    $ \widetilde{\gamma}_{i2}^{z}  \sim$  & &   &   & & $1+\text{G}(1, 2)$ & $1+\text{G}(\widetilde{\eta}_{2}^{z}, 2)$  \\ 
   $\widetilde{\gamma}_{i3}^{z}  \sim$  & &   & &  & $\text{G}(10,2.5)$ & $\text{G}(\widetilde{\eta}_{3}^{z}, 2.5)$  \\[0.3cm]   
  
 \hline
 $ \rho_{1}^{z} \sim$  & & $\text{N}(-3, 0.75)$	 & & $\text{N}(-3, 0.75)$ &    & $\text{N}(-3, 0.75)$   \\ 
  $\rho_{2}^{z} \sim$  &  & $\text{N}(0, 0.75)$  & & $\text{N}(0,0.75)$ &  &  $\text{N}(0,0.75)$  \\ 
  \vdots &   &  \vdots  &   &  \vdots &   &    \vdots    \\ 
  $ \rho_{20}^{z} \sim$    & & $\text{N}(0, 0.75)$&   &  $\text{N}(0,0.75)$ &  & $\text{N}(0,0.75)$    \\[0.3cm]

  $\eta_{1}^{z} \sim $ & &   & & $\text{G}(50, 10)  $&  & $\text{G}(50, 10)  $   \\ 
  $\eta_{2}^{z} \sim $&    & & & $\text{G}(10, 10)  $&   & $\text{G}(10, 10)  $ \\ 
  $\eta_{3}^{z} \sim $&   &  & & $\text{G}(500, 50) $ &  & $\text{G}(500, 50) $    \\[0.3cm]

 $ \widetilde{\eta}_{1}^{z} \sim$  & &  & &    &  & $\text{G}(30, 15) $    \\ 
  $\widetilde{\eta}_{2}^{z} \sim$  & &  & &    &   & $\text{G}(10, 10)  $ \\ 
  $\widetilde{\eta}_{3}^{z} \sim$  & &  & &    &   & $\text{G}(500, 50) $  \\ 
  
   \hline
  \end{tabular}
  
\end{table*}

\begin{table*}[h]\label{count_prior_table}
\caption{Prior formulation of models \ref{COUNT_NM1}, \ref{COUNT_HB}, \ref{COUNT_BE} and \ref{COUNT_HBE}.  }
\label{table:COUNT_prior_formulations}
\begin{tabular}{ccccc}
\hline
Parameter & \ref{COUNT_NM1} &  \ref{COUNT_HB} &   \ref{COUNT_BE} & \ref{COUNT_HBE}   \\
\hline
$\varphi_{i}^{c} \sim $ & $\text{N}(-4,4) $ & & &       \\ 
$\theta^c_{i1} \sim $ & & $\text{N}(\mu_{1}^{c}, 1) $ & $\text{N}(1, 0.75)$  & $\text{N}(\mu_{1}^{c}, 0.05) $      \\ 
$ \theta^c_{i2} \sim$ & & $\text{N}(\mu_{2}^{c}, 1)$  & $\text{N}(-1, 0.75)$  & $\text{N}(\mu_{2}^{c}, 0.05)$   \\[0.3cm]   
  $ \gamma_{i1}^{c} \sim $& &   & $\text{G}(1, 5)$ & $\text{G}(\eta_{1}^{c}, 5)$  \\ 
  $ \gamma_{i2}^{c} \sim $& &   & 1+$\text{G}(3, 1)$ & 1+$\text{G}(\eta_{2}^{c}, 1)$  \\ 
   $\gamma_{i3}^{c} \sim$ & &  & $\text{G}(4,1)$ & $\text{G}(\eta_{3}^{c}, 1)$    \\[0.3cm]   
\hline
$ \rho_{1}^{c} \sim$ & & $\text{N}(1, 0.5)$	  &  &  $\text{N}(1, 0.75)$     \\ 
  $\rho_{2}^{c} \sim$ &  & $\text{N}(-1, 0.5)$  &    & $\text{N}(-1, 0.75)$      \\[0.3cm]           
  $\eta_{1}^{c} \sim $& &    &  & $\text{G}(5, 5)$       \\ 
  $\eta_{2}^{c} \sim $& &    &    & $\text{G}(15, 5)$    \\ 
  $\eta_{3}^{c} \sim $& &    &    &$\text{G}(40, 10)$   \\       
   \hline
  \end{tabular}  
\end{table*}

\section*{Acknowledgements}
This work has been carried out with the financial support of the EPSRC, the Alan Turing Institute and dunnhumby ltd, our industrial partner. Access to anonymised electronic point of sale data granted by our industrial partner dunnhumby ltd.

\bibliographystyle{plainnat}

\bibliography{../../../../Bibliography_thesis.bib}

\end{document}